\documentclass[prb, twocolumn, longbibliography]{revtex4-2}
\usepackage{graphicx}
\usepackage{amsmath}
\usepackage{amssymb}
\usepackage{color}
\usepackage{appendix}
\usepackage{braket}
\usepackage{bbm}
\usepackage{subfigure}
\usepackage{mathrsfs}

\usepackage[colorlinks,urlcolor=blue,citecolor=blue,linkcolor=blue]{hyperref}
\usepackage{lipsum}
\usepackage{diagbox}
\usepackage{multirow}

\begin{document}
\title{Hybrid scale-free skin  effect in non-Hermitian systems: A transfer matrix approach}
\author{Yongxu Fu}
\email{yongxufu@pku.edu.cn}
\affiliation{International Center for Quantum Materials, School of Physics, Peking University, Beijing 100871, China}
\author{Yi Zhang}
\email{frankzhangyi@gmail.com}
\affiliation{International Center for Quantum Materials, School of Physics, Peking University, Beijing 100871, China}

\begin{abstract}
Surpassing the individual characteristics of the non-Hermitian skin effect (NHSE) and the scale-free (SF) effect observed recently, we systematically exploit the exponential decay behavior of bulk eigenstates via the transfer matrix approach in non-Hermitian systems. We concentrate on one-dimensional (1D) finite-size non-Hermitian systems with $2\times2$ transfer matrices in either the absence or presence of the boundary impurity. We analytically unveil that the unidirectional SF effect emerges with the singular transfer matrices, while the hybrid scale-free skin (SFS) effect appears with the nonsingular transfer matrices even when an open boundary condition (OBC) is imposed. The unidirectional SF effect exceeds the scope of the SF effect in previous works, while the hybrid SFS effect is an interesting interplay between the skin effect and the SF effect in finite-size systems. Our results reveal that the skin effect under the OBC prevails when it coexists with the SF effect as the system approaches the thermodynamic limit in the presence of the hybrid SFS effect. Our approach paves the way for rigorous and unified explorations of the skin and SF effects in both Hermitian and non-Hermitian systems with generic boundary conditions. 
\end{abstract}
\maketitle

\section{Introduction}
\label{section1}
Exceeding the requirement of Hermitian operators for physical observables in quantum mechanics, non-Hermitian physics has been widely broadened in the past few years \cite{gong2018,kawabataprx,ashida2020,bergholtzrev2021} to contain basic energy band theory \cite{lee2016skin,leykam2017edge,shen2018,kunst2018,yao2018,song2019,yokomizo2019,lee2019an,longhi2019,zhang2020,origin2020,yang2020,lin2023nhse,yao201802,slager2020,xue2021simple,edgeburst2022,longhi2022,fu2023ana}, higher-order topological phases \cite{kawabata2019second,lee2019ho,edvardsson2019,kawabatahigher,okugawa2020,fu2021,yu2021ho,palacios2021,st2022}, unique exceptional points (EPs) in non-Hermitian systems \cite{torres2018anomalous,kawabata2019,yokomizo2020,jones2020,zhang2020ep,yang2021,denner2021,fu2022,mandal2021ep,delplace2021ep,liu2021ep,marcus2021ep,ghorashi2021dirac,ghorashi2021weyl}, and other subjects in the scope of condensed matter physics. The exploration of the NHSE in 1D non-Hermitian systems, which supports the accumulation of an extensive number of nominal bulk eigenstates at the boundaries under the OBC on the single-particle level, is a milestone in research on non-Hermitian systems \cite{yao2018,song2019,yokomizo2019,lee2019an,longhi2019,zhang2020,origin2020,yang2020,lin2023nhse}. The non-Bloch band theory with the concept of the generalized Brillouin zone (GBZ) is fully explanatory for the NHSE \cite{yao2018,yokomizo2019,yang2020}, and more recently, similar counterparts have paved our way towards higher dimensions \cite{zhang2022uni,wang2022amoeba,yokomizo2022nonbloch,hu2023nonhermitian}. 

A typical characteristic of the NHSE eigenstates is their exponential decay in the bulk region. The exponential factor, corresponding to the relevant point on the GBZ, changes the wave vector in the traditional Bloch band theory from a real value to a complex value, a core context of the non-Bloch band theory. Strictly speaking, the complexity of the non-Hermitian systems does not constrain the localized behavior of the bulk eigenstates in the NHSE. The SF effect, which suggests the existence of eigenstates with a size-dependent localization length in 1D non-Hermitian systems, was explored recently \cite{li2020critical,yokomizo2021scaling,li2021impurity,guo2023scale,libo2023scale,molignini2023anomalous,wang2023scalefree}. Nevertheless, the SF behavior is model dependent, which is a seemingly accidental phenomenon closely related to the impurities \cite{zhu2014pt,felix2018topo,liu2020diag,takato2020impurity,liu2020exact,koch2020bbc,guo2021exact,wu2023effective}, and lacks a unified perspective in non-Hermitian systems. Furthermore, the emergence of the interplay between the skin effect and the SF effect remains an unsurveyed subject in the field of non-Hermitian physics. 

The transfer matrix approach has been a powerful tool for elucidating tight-binding models for decades \cite{transfer1,transfer2,transfer3,transfer4,transfer5,transfer6,vatsal2016transfer}, has a briefer and more compact form than the method of directly solving the eigenvalue problem, and has also been applied to non-Hermitian systems in recent years \cite{kunst2019transfer,luo2021transfer,hamed2023transfer}. Several topological properties, ranging from the topological invariants on the Riemann surface to characteristic edge states, wew also established from a transfer matrix perspective in many previous works  \cite{transfer3,transfer4,vatsal2016transfer,kunst2019transfer}. In this paper, we utilize the transfer matrix approach to establish a unified depiction of the SF effect as well as the interplay between the skin effect and the SF effect. Without loss of generality, we concentrate on 1D finite-size non-Hermitian systems with $2\times 2$ transfer matrices, such as the Hatano-Nelson (HN) model \cite{hatano1996,hatano1997,hatano1998} and the non-Hermitian Su-Schrieffer-Heeger (NH-SSH) model \cite{lee2016skin,yao2018} with or without the boundary impurity. We unveil that the unidirectional SF effect emerges with the singular transfer matrices, while the hybrid SFS effect appears with the nonsingular transfer matrices, which shows the existence of bulk eigenstates possessing both the skin effect and the SF decay factors. The unidirectional SF effect usually accompanies the boundary impurity, while the hybrid SFS effect may exist even under the OBC. The localization length of the unidirectional SF mode is generally quasilinearly dependent on the system size, which is the generalized definition of the SF effect in this paper---the results in previous works \cite{li2021impurity,guo2023scale,libo2023scale,molignini2023anomalous} are equivalent to specific cases with linear dependence. The hybrid SFS effect implies that the skin effect under the OBC may coexist with the SF effect at finite size yet prevails in the thermodynamic limit, revealing the NHSE's dominance. Once we include the boundary impurity, the hybrid SFS effect displays a finite-size dependence. Our studies complement the finite-size manifestation and skin effect localization of single-particle eigenstates in non-Hermitian systems, which may alter the (directional) visibility and transparency behaviors in photonic \cite{optical3,optical4,optical5,lin2022mobilityedge,optical6}, acoustic \cite{acoustic1,acoustic2,acoustic3}, and mechanical devices \cite{mechanical2,mechanical1,mechanical3}. 

The rest of this paper is organized as follows. In Sec.~\ref{section2}, we establish the formalism for probing unidirectional SF and hybrid SFS effects from the transfer matrix perspective, corresponding to the cases with singular and nonsingular transfer matrices, respectively. In Secs. \ref{section3a} and \ref{section4a}, we analytically solve the eigenstates as well as the energy spectrum with the emergence of the pure SF effect for the HN model and the NH-SSH model with the boundary impurity. In contrast, we establish the hybrid SFS effect in the HN model with the boundary impurity in Sec. \ref{section3b} and the NH-SSH model under the OBC in Sec. \ref{section4b}. We give a conclusion and further discussion in Sec. \ref{section5}.

\section{Probing unidirectional SF and hybrid SFS effects via the transfer matrix approach}
\label{section2}

\subsection{Review of the transfer matrix approach for non-Hermitian tight-binding models}
\label{section2a}

Commonly, the tight-binding Hamiltonian of a 1D non-interacting non-Hermitian system reads
\begin{align}
    \label{eqgeneraltba}
    \hat{\mathcal{H}}&=\sum_{n}\sum_{l=-R}^{R}\sum_{\mu,\nu=1}^{q}t_{l,\mu\nu}\hat{c}_{n,\mu}^{\dagger}\hat{c}_{n+l,\nu}\nonumber\\
    &=\sum_{n}\sum_{l=-R}^{R}\hat{\mathrm{c}}_{n}^{\dagger}\mathrm{t}_{l}\hat{\mathrm{c}}_{n+l},
\end{align}
where $\hat{c}_{n,\mu}^{\dagger}$ ($\hat{c}_{n,\mu}$) is a creation (annihilation) operator with internal degrees of freedom $\mu$ in the $n$th site, $\hat{\mathrm{c}}_{n}^{\dagger}$ ($\hat{\mathrm{c}}_{n}$) is a row (column) vector containing $q$ $\hat{c}_{n,\mu}^{\dagger}$ ($\hat{c}_{n,\mu}$), and $t_{l,\mu\nu}$ ($\mathrm{t}_{l}$) is a hopping amplitude (matrix) to the $l$th nearest site. 

Following the formalism established in Refs.~\cite{transfer1,vatsal2016transfer,kunst2019transfer}, we bundle at least $R$ adjacent sites into a supercell, such that Eq. (\ref{eqgeneraltba}) reduces to a nearest-neighbor tight-binding Hamiltonian:
\begin{align}
    \label{eqreduceham}
    \hat{\mathcal{H}}=\sum_{n=1}^{N-1}\left[\hat{\boldsymbol{c}}_{n}^{\dagger}\boldsymbol{J}_{L}\hat{\boldsymbol{c}}_{n+1}+\hat{\boldsymbol{c}}_{n}^{\dagger}\boldsymbol{M}\hat{\boldsymbol{c}}_{n}+\hat{\boldsymbol{c}}_{n+1}^{\dagger}\boldsymbol{J}_{R}^{\dagger}\hat{\boldsymbol{c}}_{n}\right],
\end{align}
with the creation (annihilation) operator $\hat{\boldsymbol{c}}_{n}^{\dagger}$ ($\hat{\boldsymbol{c}}_{n}$) for $N$ supercells. There are $\mathcal{N}\geq qR$ internal degrees of freedom in each supercell, and $\boldsymbol{J}_{L, R}$ and $\boldsymbol{M}$ are the hopping matrices between neighboring supercells and the on-site matrix, respectively. Without loss of generality, we introduce non-Hermiticity on only $\boldsymbol{M}$, i.e., $\boldsymbol{M}\neq \boldsymbol{M}^{\dagger}$, and set $\boldsymbol{J}_{R}=\boldsymbol{J}_{L}\equiv \boldsymbol{J}$, with $\boldsymbol{J}^{2}=0$ \footnote{This setup is adequate for the core physics in the current paper; also, we can always ensure the nilpotence of $\boldsymbol{J}$ by bundling a sufficient number of unit cells. }. Consequently, the tight-binding Hamiltonian further reduces to
\begin{align}
    \label{eqsimplereduceham}
    \hat{\mathcal{H}}=\sum_{n=1}^{N-1}\left[\hat{\boldsymbol{c}}_{n}^{\dagger}\boldsymbol{J}\hat{\boldsymbol{c}}_{n+1}+\hat{\boldsymbol{c}}_{n}^{\dagger}\boldsymbol{M}\hat{\boldsymbol{c}}_{n}+\hat{\boldsymbol{c}}_{n+1}^{\dagger}\boldsymbol{J}^{\dagger}\hat{\boldsymbol{c}}_{n}\right].
\end{align}  
Given an arbitrary single-particle state
\begin{align}
    \label{eqgeneralstateform}
    \ket{\Psi}=\sum_{n=1}^{N}\Psi_{n}\hat{\boldsymbol{c}}_{n}^{\dagger}\ket{0},
\end{align}
with $\Psi_{n} \in \mathbb{C}^{\mathcal{N}}$, the single-particle Schr\"odinger equation $\hat{\mathcal{H}}\ket{\Psi}=\varepsilon\ket{\Psi}$ reduces to the recursion relation
\begin{align}
    \label{eqrecursionrelation}
    \boldsymbol{J}\Psi_{n+1}+\boldsymbol{J}^{\dagger}\Psi_{n-1}=\left(\varepsilon\mathbbm{1}-\boldsymbol{M}\right)\Psi_{n}.
\end{align}

Next, we define $\mathcal{G}=\left(\varepsilon\mathbbm{1}-\boldsymbol{M}\right)^{-1}$ as the on-site Green's function, which is nonsingular except when $\varepsilon$ is an eigenvalue of $\boldsymbol{M}$. Performing the reduced singular value decomposition (SVD) \cite{lay2016,strang2023}, we obtain
\begin{align}
    \label{eqsvd}
    \boldsymbol{J}=V\Xi W^{\dagger},
\end{align}
where $\Xi=\text{diag}\left\{\xi_{1},\ldots,\xi_{r}\right\}$ is a diagonal matrix of singular value $\xi_{i}\in\mathbb{R}^{+}, i=1,2,\ldots,r$, with $r=\text{rank}\left(\boldsymbol{J}\right)$, and $V$~($W^{\dagger}$) is the matrix composed of $r$ orthonormal bases $v_{i}$~($w_{i}^{\dagger}$) in the column~(row) space of $\boldsymbol{J}$, that is,
\begin{align}
    \label{orthogonalrelation}
    V^{\dagger}V=W^{\dagger}W=\mathbbm{1},\qquad V^{\dagger}W=0.
\end{align}

For simplicity, we focus on the $r=1$ and $\mathcal{N}=2$ case for demonstrations; as a result, $\Xi\equiv\xi\in\mathbb{R}^{+}$, and $\left\{V\equiv v, W\equiv w\right\}$ constitutes a set of orthonormal bases of $\mathbb{C}^{2}$. In this basis, we expand $\Psi_{n}$ and $\mathcal{G}$ as  
\begin{align}
    \label{eqwavebasis}
    \Psi_{n}=\alpha_{n}v+\beta_{n}w,\quad
    \alpha_{n}=v^{\dagger}\Psi_{n},\quad \beta_{n}=w^{\dagger}\Psi_{n},
\end{align}
and
\begin{align}
    \label{eqonsitgreenexp}
    \mathcal{G}_{AB}=B^{\dagger}\mathcal{G}A\in\mathbb{C},\qquad A,B\in \left\{v, w\right\},
\end{align}
respectively.
Combining Eqs. (\ref{eqrecursionrelation})-(\ref{eqonsitgreenexp}), we obtain the propagating relation in the bulk
\begin{align}
    \label{eqtransfereq}
    \Phi_{n+1}=T\Phi_{n},\qquad 
    \Phi_{n}\equiv\left(\begin{matrix}
        \beta_{n}\\
        \alpha_{n-1}
    \end{matrix}\right),
\end{align}
where $T$ is the $2\times 2$ transfer matrix \cite{kunst2019transfer}:
\begin{align}
    \label{eqtransfermatrix}
    T=\frac{1}{\xi\mathcal{G}_{vw}}\left(\begin{matrix}
        1&-\xi\mathcal{G}_{ww}\\
        \xi\mathcal{G}_{vv}&\xi^{2}\left(\mathcal{G}_{vw}\mathcal{G}_{wv}-\mathcal{G}_{vv}\mathcal{G}_{ww}\right)
    \end{matrix}\right).
\end{align}

We define the trace and determinant of $T$ as 
\begin{align}
    \label{eqtracedet}
    \Delta=\text{tr}\left(T\right),\qquad \Gamma=\det{\left(T\right)}\equiv\frac{\mathcal{G}_{wv}}{\mathcal{G}_{vw}},
\end{align}
which are rational functions of the energy $\varepsilon$ in general; hereafter, we suppress their explicit dependence on $\varepsilon$ for simplicity. If $T$ is singular ($\Gamma=0$), $T^{n}=\Delta^{n-1}T$; otherwise, $T$ is nonsingular ($\Gamma\neq0$) \cite{kunst2019transfer}, and  \begin{align}
    \label{eqnonsingtn}  
    T^{n}=\Gamma^{n/2}\left[\frac{U_{n-1}(z)}{\sqrt{\Gamma}}T-U_{n-2}(z)\mathbbm{1}\right],
\end{align}
where
\begin{align}
    \label{eqchebyshev} 
    U_{n}(z)=\frac{\sin{\left[\left(n+1\right)\phi\right]}}{\sin{\phi}}
\end{align}  
is the Chebyshev polynomials of the second kind \cite{chebyshevpoly1,chebyshevpoly2} and
\begin{align}
    \label{eqzequalcos}
    z\equiv z(\varepsilon)=\frac{\Delta}{2\sqrt{\Gamma}}=\cos{\phi}\in\mathbb{C}.
\end{align}

After reviewing the transfer matrix approach for non-Hermitian tight-binding models, we shall utilize it to probe unidirectional SF and hybrid SFS effects. 

\begin{figure} 
    \centering 
    \subfigure[]{\includegraphics[width=8.5cm, height=2.8cm]{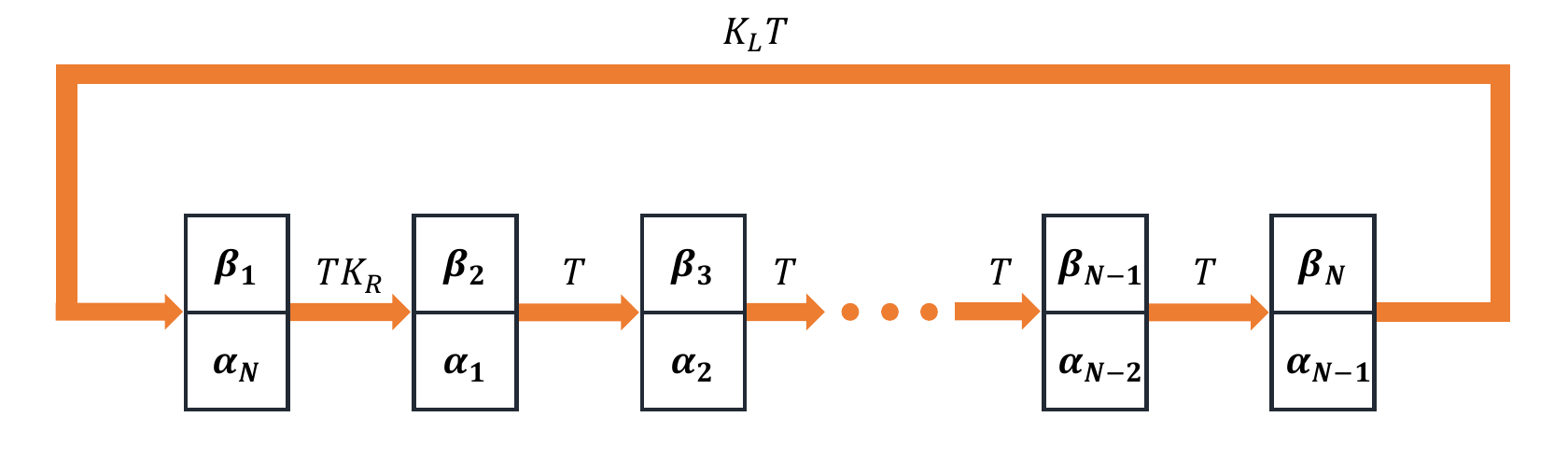}}
    \subfigure[]{\includegraphics[width=7.5cm, height=2cm]{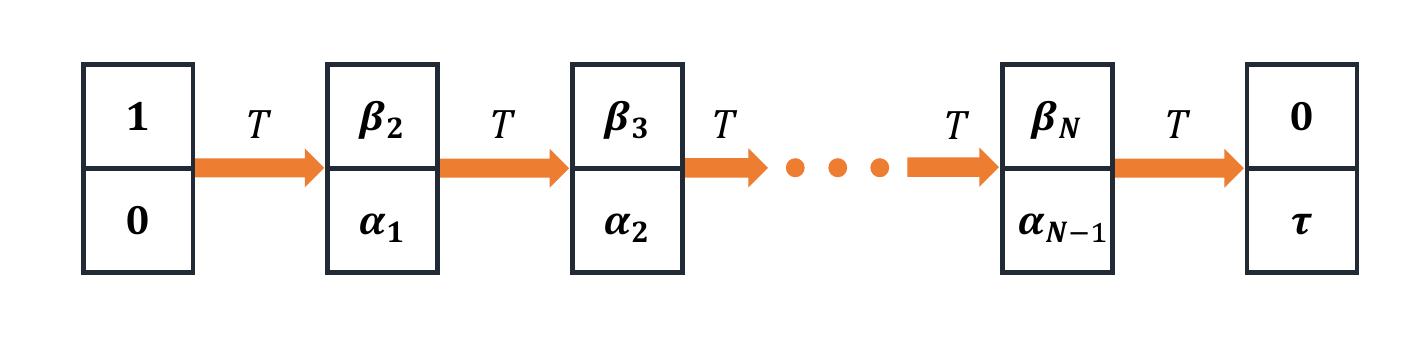}}
    \caption{The propagating relation of $\Phi_{n}$ in the transfer matrix approach (a) with the boundary impurity in Eq. (\ref{eqboundimp}) and (b) under the OBC as in Eqs. (\ref{eqtransfereq}) and (\ref{eqgeneralbc}) helps us establish the energy spectrum and the corresponding eigenstates. }
    \label{Fig-propageting}
\end{figure}

\subsection{The transfer matrix approach for 1D tight-binding models with boundary impurities}
\label{section2b}

Without loss of generality, in addition to the Hamiltonian in Eq.~(\ref{eqsimplereduceham}), we include the boundary impurity term  
\begin{align}
    \label{eqboundimp}   \hat{\boldsymbol{c}}_{N}^{\dagger}\kappa_{L}\hat{\boldsymbol{c}}_{1}+\hat{\boldsymbol{c}}_{1}^{\dagger}\kappa_{R}\hat{\boldsymbol{c}}_{N}, 
\end{align}
with corresponding hopping matrices $\kappa_{L}=\gamma_{L}\boldsymbol{J}$ and $\kappa_{R}=\gamma_{R}\boldsymbol{J}^{\dagger}$, with $\gamma_{L},\gamma_{R}\geq0$, connecting the first and last supercells \footnote{More general boundary impurities are still solvable within the current transfer matrix framework with increased supercell sizes and thus computational complexity. }. Note that the periodic boundary condition, as well as the Bloch band theory, is recovered in the $\gamma_{L}=\gamma_{R}=1$ case, while the OBC corresponds to the $\gamma_{L}=\gamma_{R}=0$ case. After some algebra, we obtain the following physical condition on the transfer matrix (Appendix \ref{appendixa}):
\begin{align}
    \label{eqgeneralbc}
    \Phi_{1}=K_{L} T\Phi_{N},\qquad \Phi_{2}=TK_{R}\Phi_{1},
\end{align}
where $K_{L}=\text{diag}\left\{1/\gamma_{L},1\right\}$ and $K_{R}=\text{diag}\left\{1,\gamma_{R}\right\}$. Together with Eq. (\ref{eqtransfereq}), we illustrate the full propagating relation of $\Phi_{n}$ with the boundary impurity in Fig. \ref{Fig-propageting}(a). Since $\Phi_{N}=T^{N-2}\Phi_{2}$, Eq. (\ref{eqgeneralbc}) gives
\begin{align}
    \label{eqboundbc}
    \Phi_{1}=K_{L}T^{N}K_{R}\Phi_{1}.
\end{align}
Defining $\varphi=K_{R}\Phi_{1}$ and $K=K_{R}K_{L}=\text{diag}\left\{1/\gamma_{L},\gamma_{R}\right\}$, we finally obtain the consistency equation with the boundary conditions
\begin{align}
    \label{eqgeneralbeq}
    \varphi=KT^{N}\varphi,
\end{align}
which implies that the legitimate $\varphi$ must be the eigenvector of $KT^{N}$ with an eigenvalue $1$ and, subsequently, $\Phi_{1}=K_{R}^{-1}\varphi$ and $\Phi_{n}=T^{n-1}\varphi$ for $n=2,3,\ldots,N$.

\subsubsection{The pure SF effect}
\label{section2b1}

We first consider the cases with singular transfer matrices, which correspond to the occurrence of real-space EPs under the OBC \cite{kunst2019transfer}. As we turn on the boundary impurity (\ref{eqboundimp}), the determinant of $KT^{N}$ vanishes because $\Gamma=0$, which, together with Eq. (\ref{eqgeneralbeq}), implies $\text{tr}\left(KT^{N}\right)=1$. Consequently, $\text{tr}\left(KT\right)\neq0$ under normal circumstances; we obtain $\Delta^{N-1}\text{tr}\left(KT\right)=1$, namely,
\begin{align}
    \label{eqgeneralpuresf}
    \Delta^{N}=\frac{\Delta}{\text{tr}\left(KT\right)}.
\end{align}
Defining a constant $c=\Delta^{-N}$, we can express $\Delta$ as
\begin{align}
    \label{eqgeneralpuresfcal}
    \Delta=c^{-\frac{1}{N}}e^{-i\frac{2\pi m}{N}}
\end{align}
for each $m\in\left\{1,2,\ldots,N\right\}$. Consequently, putting together Eq. (\ref{eqgeneralpuresf}) and $\Delta$'s energy $\varepsilon$ dependence, we can solve for $c_m$, $\Delta_m$, and the energy $\varepsilon_m$ for each $m$ (in addition to the potential band index) \footnote{We may encounter multiple solutions or no solution for certain $m$ values. }, generating the nominal bulk bands. 

Consider a fixed $c_{m}$; the eigenstate is given by
\begin{align}
    \label{eqsoluscale}
    \Phi_{1}^{m}&=K_{R}^{-1}\varphi_{m}, \nonumber\\
    \Phi_{n}^{m}&=c_{m}^{-\frac{n-2}{N}}e^{-i\frac{2\pi m}{N}(n-2)}T\varphi_{m},
\end{align}
where $n=2,3,\ldots,N$ and $KT^{N}\varphi_{m}=\varphi_{m}$. According to Eq. (\ref{eqwavebasis}), we obtain the eigenstate (up to normalization here and afterward) corresponding to energy $\varepsilon_{m}$
\begin{align}       \Psi_{1}^{m}&=\left(T\varphi_{m}\right)_{2}v+\left(K_{R}^{-1}\varphi_{m}\right)_{1}w,\label{eqeigenstatescale1}\\
    \Psi_{n}^{m}&=c_{m}^{-\frac{n-1}{N}}e^{-i\frac{2\pi m}{N}(n-1)}\psi_{m},\,\, n=2,3,\ldots,N-1,\label{eqeigenstatescalebulk}\\
    \Psi_{N}^{m}&=\left(K_{R}^{-1}\varphi_{m}\right)_{2}v+\left(c_{m}^{-\frac{N-2}{N}}e^{-i\frac{2\pi m}{N}(N-2)}T\varphi_{m}\right)_{1}w\label{eqeigenstatescalen},
\end{align}
where $\psi_{m}=\left(T\varphi_{m}\right)_{2}v+c_{m}^{\frac{1}{N}}e^{i\frac{2\pi m}{N}}\left(T\varphi_{m}\right)_{1}w$, and $\left(\mathscr{V}\right)_{1,2}$ denote the first and the second components of the column $2$-vector $\mathscr{V}$. 

We denote the bulk region as $\mathscr{B}=\left\{2,3,\ldots,N-1\right\}$ and the boundary region as $\mathscr{E}=\left\{1,N\right\}$ for the system in Eq.~(\ref{eqsimplereduceham}) with the boundary impurity in Eq.~(\ref{eqboundimp}). The bulk solution in Eq. (\ref{eqeigenstatescalebulk}) clearly exhibits an exponential decay,
\begin{align}
    \label{eqscaleexp}
      \exp{\left\{-\frac{\text{Re}(\log{c_{m}})}{N}(n-1)\right\}},
\end{align}
with a phase factor
\begin{align}
    \label{eqphasrfac} 
    \exp{\left\{-i\frac{\text{Im}(\log{c_{m}})+2\pi m}{N}(n-1)\right\}},
\end{align} 
which is a correction with wave vector $2\pi m/N$. We define the occurrence of such an exponential decay behavior of eigenstates in $\mathscr{B}$ as the unidirectional SF effect, which is a generalization of the SF effect in previous works \cite{li2020critical,yokomizo2021scaling,li2021impurity,guo2023scale,libo2023scale,molignini2023anomalous}. Further, we denote the eigenstates in Eq.~(\ref{eqeigenstatescalebulk}) with $\text{Re}(\log{c_{m}})>0$~($<0$) as the left- (right)-accumulation unidirectional SF modes, whose localization lengths are  
\begin{align}
    \label{eqlocalength}
    \xi_{m}=\frac{N}{|\text{Re}(\log{c_{m}})|}.
\end{align}
Noteworthily, $\log{c_{m}}$ is generally sublinear in the system size $N$, and $\xi_{m}$ is approximately linearly dependent on $N$, a typical character of the unidirectional SF effect. However, such quasilinear dependence of the localization length $\xi_{m}$ on $N$ vanishes in the thermodynamic limit or with sufficiently large $N$ under certain scenarios, as demonstrated in later examples and the Appendixes.

\subsubsection{The hybrid SFS effect}
\label{section2b2}

More exotic phenomena beyond the pure NHSE emerge in the case of nonsingular transfer matrices~\footnote{Here, we assume all physical energies corresponding to $\Gamma\neq0$, neglecting particular situations where specific discrete energies lead to $\Gamma=0$. }. When the system size is finite, the skin effect, terminology defined rigorously in the thermodynamic limit under the OBC, shows not only its finite-size behavior but also a potential interplay with the SF effect.  

The propagating relation of $\Phi_{n}$ under the OBC is illustrated in Fig. \ref{Fig-propageting}(b), and accordingly, the physical condition [the constraint on $\phi$ in Eq. (\ref{eqnonsingtn})] is (see Appendix \ref{appendixb})
\begin{align}
    \label{eqcondition}
    \frac{\sin{\left(N\phi\right)}}{\sin{\left[(N-1)\phi\right]}}=q,
\end{align}
with $q=\xi\sqrt{\frac{\mathcal{G}_{wv}}{\mathcal{G}_{vw}}}\mathcal{G}_{vw}$. Here, we have not simplified $q$ to avoid the branch ambiguity of the square root. Together with Eq. (\ref{eqzequalcos}), we can, in principle, obtain the energy $\varepsilon$ and $\phi$ simultaneously. If the energy $\varepsilon$ renders $q$ real, the left-hand side of Eq. (\ref{eqcondition}) must be real, thus leading to real $\phi$. However, if the energy $\varepsilon$ renders $q$ complex, we obtain a complex $\phi=\phi_{R}+i\phi_{I}$, where $\phi_{I}\sim c/N$ is quasilinearly dependent on $1/N$ (see Appendix \ref{appendixb}). The eigenstate corresponding to energy $\varepsilon$ reads
\begin{align}
    \label{eqobceigenstate}
    \Psi_{n}=\Gamma^{n/2}\left[\mathscr{A}_{L}(\phi)e^{in\phi_{R}}e^{-n\phi_{I}}+\mathscr{A}_{R}(\phi)e^{-in\phi_{R}}e^{n\phi_{I}}\right],
\end{align}
where $n=1,2,\ldots,N$ and the coefficients $\mathscr{A}_{L}(\phi)$ and $\mathscr{A}_{R}(\phi)$ are detailed in Appendix \ref{appendixb}. 

\begin{figure} 
    \centering 
    \includegraphics[width=6.5cm, height=4cm]{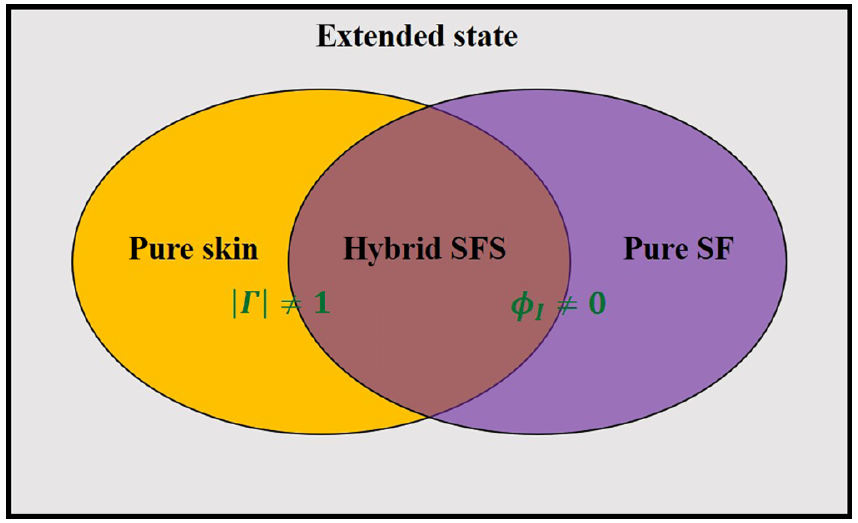}
    \caption{Schematic illustration of the relationships and differences between the conditions for an extended state, the pure skin effect, the pure SF effect, and the hybrid SFS effect with a nonsingular transfer matrix ($\Gamma\neq 0$). }
    \label{Fig-summary}
\end{figure}

The exponential factor $\Gamma^{n/2}$ indicates the emergence of the skin effect. In particular, the skin effect emerges (is absent) for $|\Gamma|\neq1$ ($|\Gamma|=1$). For instance, there is no skin effect in the Hermitian cases or the non-Hermitian cases with some special symmetries (such as parity-time (PT) symmetry \cite{kunst2019transfer}), where $|\Gamma|=1$. The additional exponential factors $e^{\pm n\phi_{I}}$ indicate the SF effect with bidirectional accumulation for a set of $(N,\phi_{R})$. Therefore, we obtain the bidirectional SF effect for $|\Gamma|=1$ and $\phi_{I}\neq0$, while $|\Gamma|=1$ and $\phi_{I}=0$ suggest an extended state. Noteworthily, the bidirectional SF effect reduces to the unidirectional SF effect if one of $\mathscr{A}_{L, R}(\phi)$ vanishes; thus, our formalism for the pure SF effect unifies the unidirectional and bidirectional SF effects. Further, the pure skin effect emerges with $|\Gamma|\neq1$ and $\phi_{I}=0$, while the SF effect and skin effect coexist for $|\Gamma|\neq1$ and $\phi_{I}\neq0$, which we define as the hybrid SFS effect. We illustrate and summarize the relationships and differences between (the conditions and properties of) the extended state, the pure skin effect, the pure SF effect, and the hybrid SFS effect in Fig. \ref{Fig-summary} and Table \ref{table1}, which generalizes and unifies the NHSE and the SF effect in previous works \cite{yao2018,song2019,yokomizo2019,li2021impurity,libo2023scale,guo2023scale,molignini2023anomalous}.  

Asymptotically, in the bulk and for $N\rightarrow +\infty$ under the OBC, however, $n\phi_{I}\rightarrow 0$, suggesting that the coexisting effects become dominated by $\Gamma^{n/2}$---the skin effect---in the thermodynamic limit \cite{fu2023ana}. Hence, the emergent SF effect may coincide with the skin effect only under finite sizes; with increasing system size, the skin effect gradually and eventually dominates. 

\begin{table*}
    \renewcommand{\arraystretch}{2.5}
    \centering
    \setlength{\tabcolsep}{1mm}{
    \begin{tabular}{c|c|c|c}
         \hline
         \hline
         $T$ & $\Gamma$ value & $\phi_{I}=0$ & $\phi_{I}\neq0$  \\
         \hline
         Singular & $\Gamma=0$ & \textbf{Extended state} & \textbf{Unidirectional SF} \\
         \hline
         \multirow{2}{*}{Nonsingular} & $|\Gamma|=1$ & \textbf{Extended state} & \textbf{Unidirectional/bidirectional SF} \\
         \cline{2-4}
         \multirow{2}{*}{} & $|\Gamma|\neq1$ & \textbf{Pure skin} & \textbf{Hybrid SFS} \\
         \hline
         \hline
    \end{tabular}}
    \caption{The conditions and behaviors of the extended state, the pure skin effect, the pure SF effect (unidirectional and bidirectional SF effects with a nonsingular transfer matrix), and the hybrid SFS effect. $\phi_{I}=\text{Re}(\log{c_{m}})/N$ in the presence of the boundary impurity in Eq. (\ref{eqboundimp}) with a singular transfer matrix.}\label{table1}
\end{table*}

Next, we include the boundary impurity in Eq. (\ref{eqboundimp}), whose physical condition is in Eq. (\ref{eqgeneralbeq}). Again, a complex $\phi=\phi_{R}+i\phi_{I}$ emerges under finite size, but the form of $\phi_{I}\sim c/N$ does not always exist when $N\rightarrow+\infty$ (see Appendix \ref{appendixc}). The general form of the eigenstate in $\mathscr{B}$ with respect to the energy $\varepsilon$ and the complex $\phi$ reads
\begin{align}
    \label{eqimpbulkeigenstate}
    \Psi_{n}^{imp}&=\Gamma^{n/2}\left[\mathcal{B}_{L}(\phi)e^{in\phi_{R}}e^{-n\phi_{I}}+\mathcal{B}_{R}(\phi)e^{-in\phi_{R}}e^{n\phi_{I}}\right],
\end{align}
where the coefficients $\mathcal{B}_{L}(\phi)$ and $\mathcal{B}_{R}(\phi)$ are given in Appendix \ref{appendixc}. Also, like for the cases under the OBC and finite size, the various conditions and behaviors of the eigenstates in the presence of the boundary impurity are depicted in Fig. \ref{Fig-summary} and Table \ref{table1}. As the skin effect is sensitive to boundary conditions and vanishes in the thermodynamic limit with boundary impurities, so is the hybrid SFS effect in the presence of boundary impurities.

\section{The Hatano-Nelson model with the boundary impurity}
\label{section3}

As a typical 1D non-Hermitian model, the HN model with the boundary impurity reads
\begin{align}
    \label{eqhaimp}
    \hat{\mathcal{H}}_{hn}=\sum_{n=1}^{N-1}\left(t_{L}\hat{c}_{n}^{\dagger}\hat{c}_{n+1}+t_{R}\hat{c}_{n+1}^{\dagger}\hat{c}_{n}\right)+\gamma_{R}\hat{c}_{1}^{\dagger}\hat{c}_{N}+\gamma_{L}\hat{c}_{N}^{\dagger}\hat{c}_{1},
\end{align}
where we assume $t_{L},t_{R},\gamma_{L},\gamma_{R}\geq0$ for simplicity. In addition to the skin effect, the model may also host the SF effect \cite{li2021impurity,molignini2023anomalous}. In this section, we apply the transfer matrix approach to comprehensively study the display and interplay of the skin effect and the SF effect in Eq. (\ref{eqhaimp}). Note that the single-band nature of the HN model offers an expedient expression of the transfer matrix in the single-particle wave function space, $\ket{\Psi}=\sum_{n=1}^{N}\psi_{n}\hat{c}_{n}^{\dagger}\ket{0}$, instead of the nominal $\left\{v,w\right\}$ space. After some algebra (Appendix \ref{appendixd}), we obtain the propagating relation in the bulk, 
\begin{align}
    \label{eqhnimpprop}
    \left(\begin{matrix}\psi_{n+1}\\ \psi_{n}
    \end{matrix}\right)=T\left(\begin{matrix}\psi_{n}\\ \psi_{n-1}
    \end{matrix}\right),
\end{align}
where the transfer matrix with respect to the energy $\varepsilon$ is 
\begin{align}
    \label{eqhntransfermat}
    T=\left(\begin{matrix}\frac{\varepsilon}{t_{L}}&-\frac{t_{R}}{t_{L}}\\1& 0
    \end{matrix}\right),
\end{align}
and we denote 
\begin{align}
    \label{eqhnimptrace}
    \Delta&=\text{tr}\left(T\right)=\frac{\varepsilon}{t_{L}},\nonumber\\
    \Gamma&=\det{\left(T\right)}=\frac{t_{R}}{t_{L}}.
\end{align}

\subsection{Exact solutions of the pure SF effect}
\label{section3a}

\begin{figure} 
    \centering 
    \subfigure{\includegraphics[width=4.1cm, height=3.3cm]{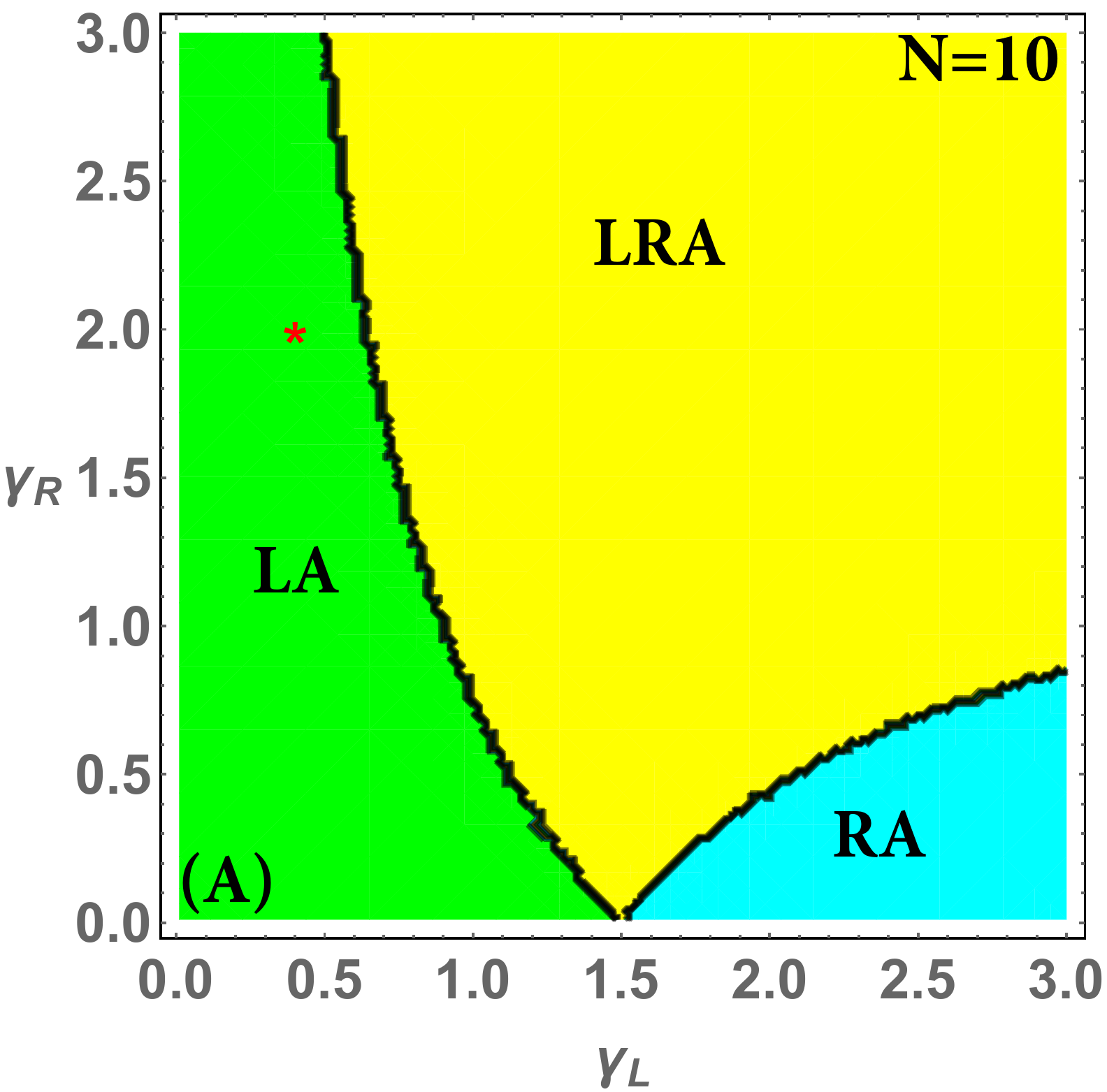}}
    \subfigure{\includegraphics[width=3.9cm, height=3.3cm]{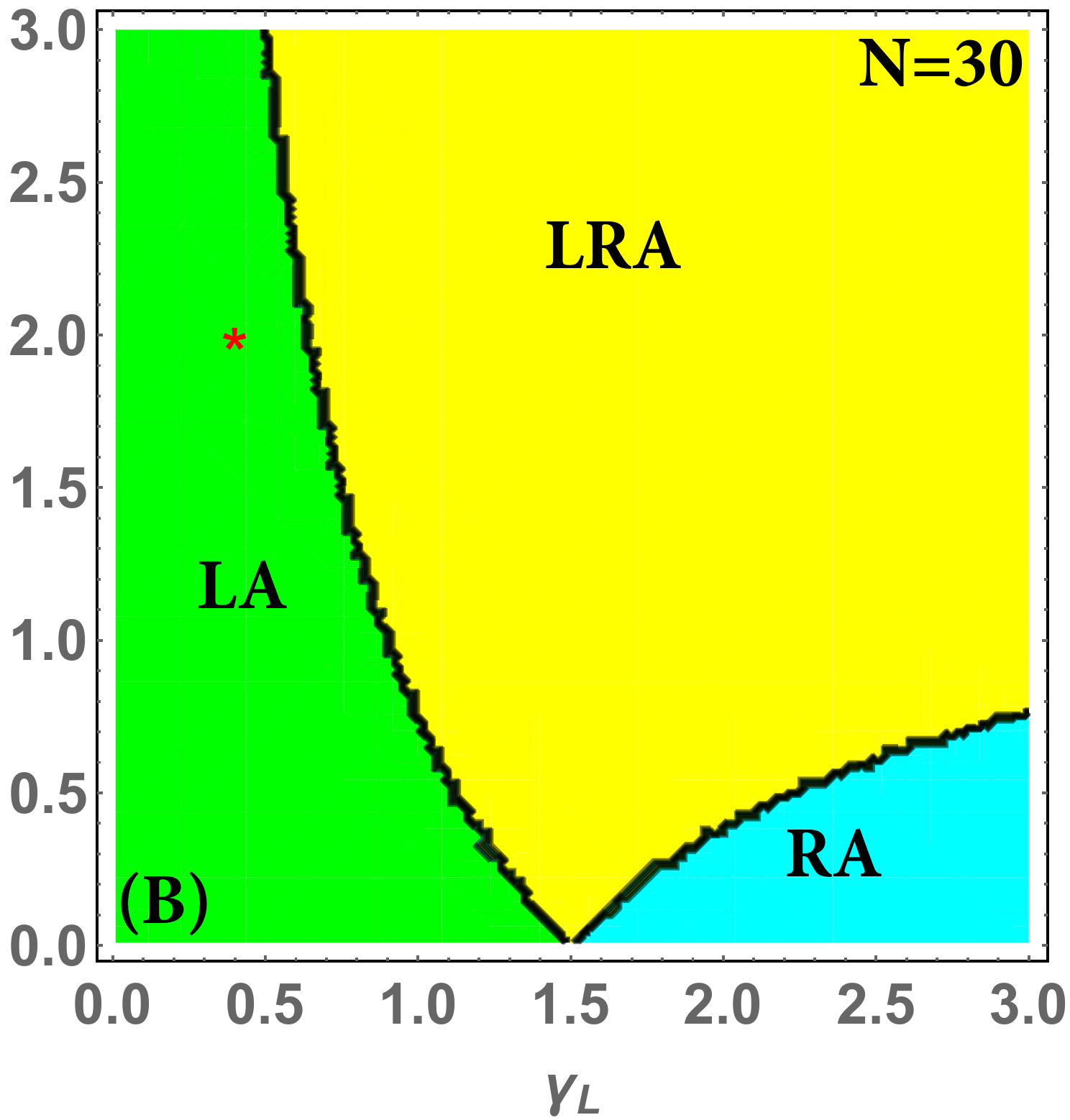}}\\
    \subfigure{\includegraphics[width=4cm, height=3cm]{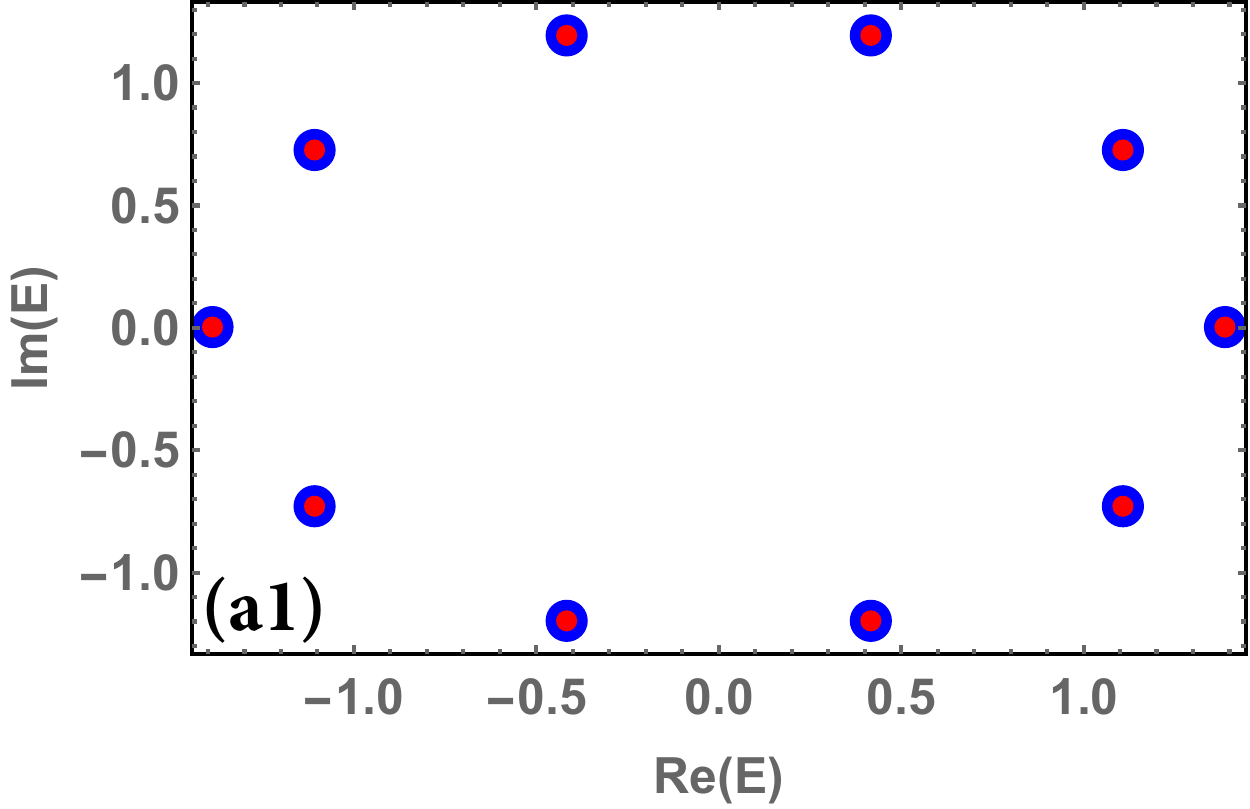}}
    \subfigure{\includegraphics[width=3.9cm, height=3cm]{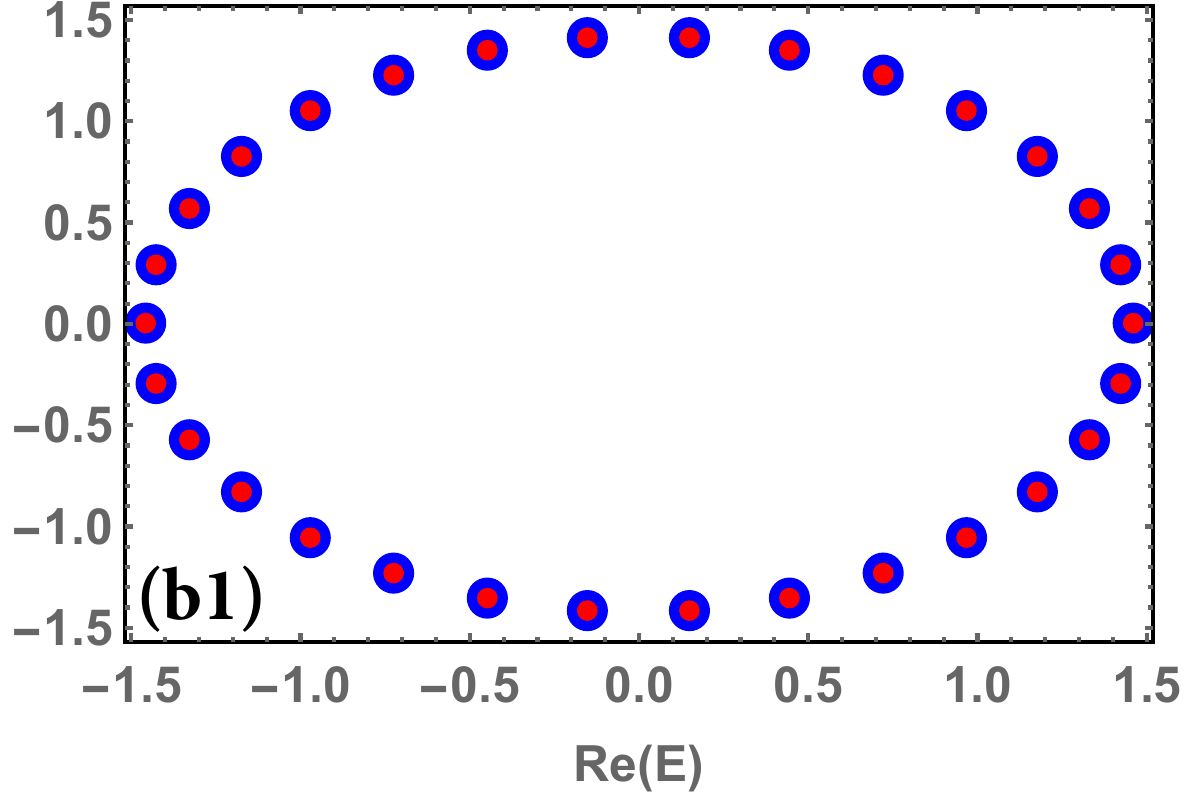}}\\
    \subfigure{\includegraphics[width=4cm, height=3cm]{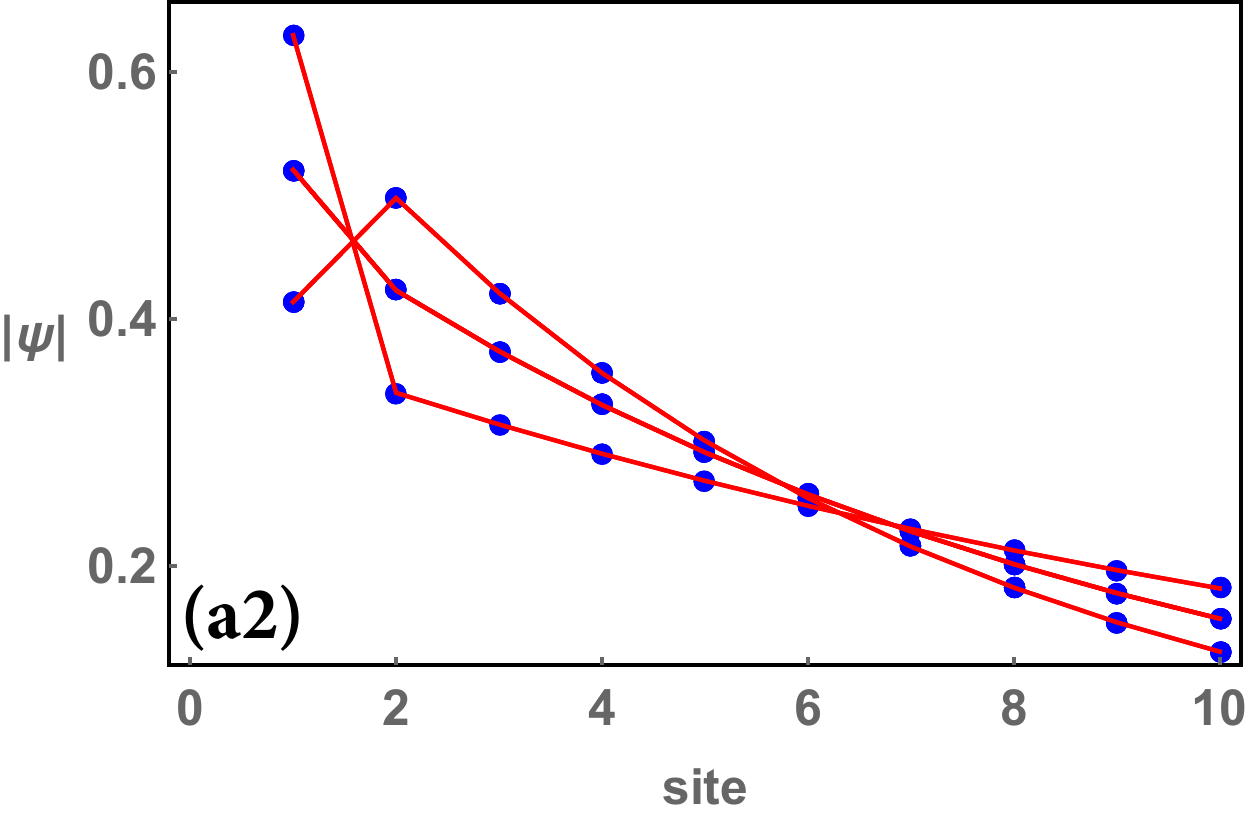}}
    \subfigure{\includegraphics[width=3.9cm, height=3.02cm]{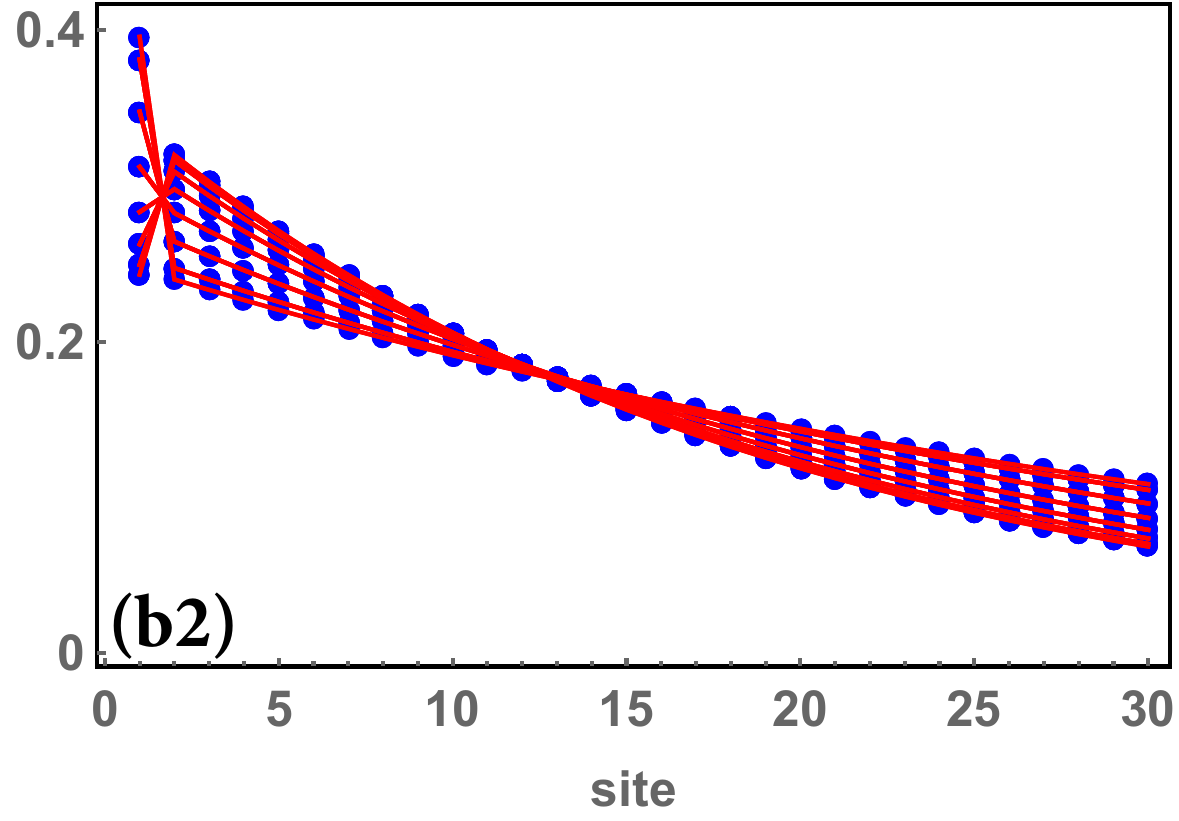}}
    \subfigure{\includegraphics[width=8cm, height=5cm]{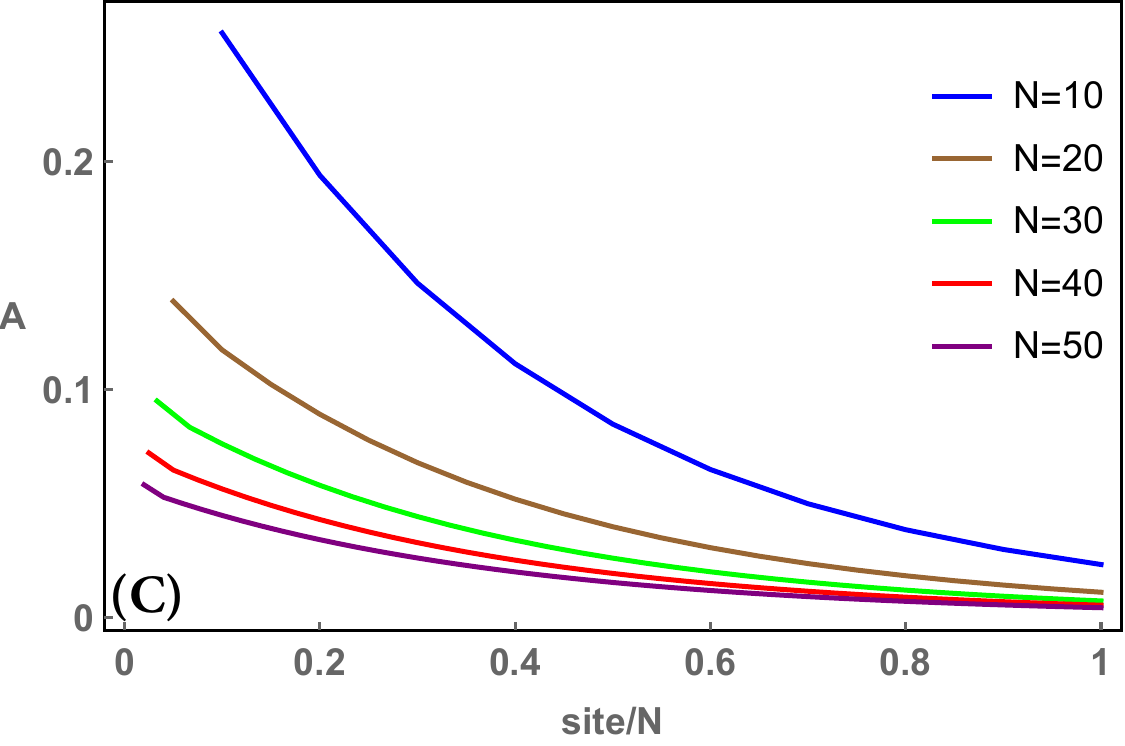}}
    \caption{The phase diagrams in the $\gamma_{L}$-$\gamma_{R}$ plane of the HN model in Eq. (\ref{eqhaimp}) for (A) $N=10$ and (B) $N=30$. The parameters $t_{L}=1.5$ and $t_{R}=0$ lead to a singular transfer matrix. (a1) and (b1) The analytic (red dots) and numerical (blue dots) results of the energy spectrum for $\gamma_{L}=0.4$ and $\gamma_{R}=2$, denoted as the red stars in (A) and (B). (a2) and (b2) The analytic (red lines) and numerical (blue dots) results of the normalized left-accumulation unidirectional SF modes corresponding to (a1) and (b1), respectively. (C) The MASEs for $t_{L}=1.5$, $t_{R}=0$, $\gamma_{L}=0.4$, and $\gamma_{R}=2$ and various $N$ display a quasilinearly dependent localization length. }
    \label{Fig-hnsingleft}
\end{figure}

First, we consider the case with a singular transfer matrix, i.e., $t_{R}=0$ that leads to $\Gamma=0$. The eigenvectors and the corresponding energies are (Appendix \ref{appendixd1})
\begin{align}
    \label{eqhnsingvec} 
    \varepsilon_{m}&=t_{L}c_{m}^{-\frac{1}{N-2}}e^{-i\frac{2\pi m}{N-2}},\quad m=1,2,\ldots,N,\nonumber\\ 
    \psi_{n}^{m}&=c_{m}^{-\frac{n-2}{N-2}}e^{-i\frac{2\pi m}{N-2}(n-2)},\quad n=3,4,\ldots,N, \nonumber\\
    \psi_{1}^{m}&=\frac{t_{L}}{\gamma_{L}}c_{m}^{-\frac{1}{N-2}}e^{-i\frac{2\pi m}{N-2}}\psi_{N}^{m},    
\end{align}
with the physical condition
\begin{align}
    \label{eqhnimpphycond3}
    \frac{t_{L}^{2}}{\gamma_{L}}c_{m}^{-\frac{2}{N-2}}e^{-i\frac{4\pi m}{N-2}}=t_{L}c_{m}+\gamma_{R},
\end{align}
where we have set $\psi_{2}^{m}=1$. Apparently, the exponential factor of $\psi_{n}^{m}$ leads to the pure SF effect, with the localization length 
\begin{align}
    \label{eqhnsingloclen}
    \xi_{m}=\frac{N-2}{|\text{Re}\left(\log{c_{m}}\right)|},
\end{align}
quasilinearly dependent on the system size, and the left- (right)-accumulation unidirectional SF modes correspond to $\text{Re}\left(\log{c_{m}}\right)>0$ ($<0$). We plot the phase diagrams of the unidirectional SF modes in the $\gamma_{L}$-$\gamma_{R}$ plane in Figs.~\ref{Fig-hnsingleft}(A)(B) with system sizes $N=10, 30$, where the green regions label the existence of the unidirectional SF modes with left-accumulation (LA), the cyan regions label that with right-accumulation (RA), and the yellow regions label that with both left- and right-accumulation (LRA) \footnote{The LRA region may contain the extreme modes with $\text{Re}\left(\log{c_{m}}\right)=0$, which are beyond the scope of the current paper. }. In Fig.~\ref{Fig-hnsingleft}, we also plot the full single-particle energy spectrum [Figs.~\ref{Fig-hnsingleft}(a1)(b1)] with the corresponding left-accumulation unidirectional SF modes [Figs.~\ref{Fig-hnsingleft}(a2)(b2)] for typical parameters [red stars in Figs.~\ref{Fig-hnsingleft}(A)(B)], where the analytic results Eq. (\ref{eqhnsingvec}) (red dots and lines) are perfectly consistent with the numerical results (blue dots) for various system sizes \footnote{We uniformly label the coordinate axis of the numerical energy spectrum (eigenstates) as $(\text{Re}(\text{E}),\text{Im}(\text{E}))$ [($\text{site},|\psi|$)] for simplicity throughout this paper. }. We define the mean amplitude squared of all eigenvectors (MASE) as
\begin{equation}
    \label{eqmeanamp}
    A(n)=\frac{1}{N}\sum_{m=1}^{N}|\psi_{n}^{m}|^{2},
\end{equation}
which we illustrate for various system sizes in Fig. \ref{Fig-hnsingleft}(C), implying the quasilinear dependence---the localization length increases with the systems size. In addition, the localized SF modes given in Refs. \cite{li2021impurity,molignini2023anomalous} are equivalent to the strong non-reciprocity or extremely special $\gamma_{R}\rightarrow0$ cases in our transfer matrix formalism~(Appendix \ref{appendixd1}). We emphasize that the quasilinear dependence of the pure SF effect is fragile for large system sizes, which may gradually evolve into the pure skin effect in the thermodynamic limit (see Appendix \ref{appendixadd1} for details).

\begin{figure*} 
    \centering 
    \subfigure{\includegraphics[width=5cm, height=3.5cm]{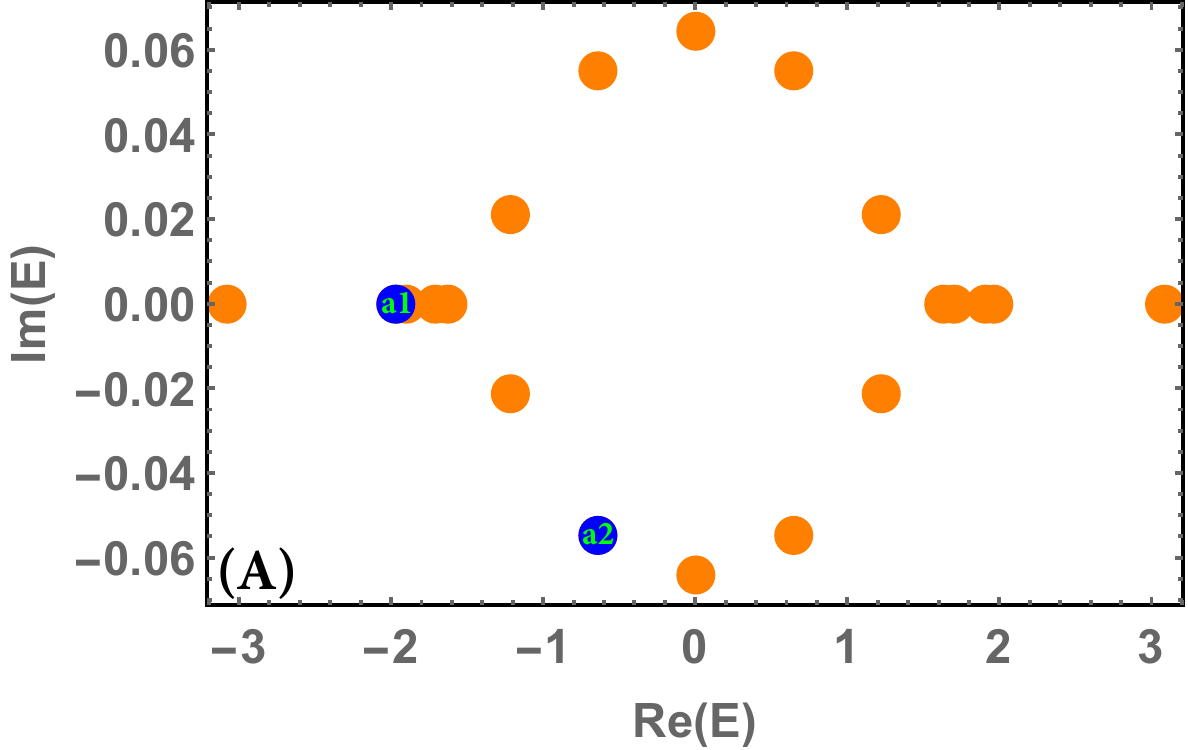}}
    \subfigure{\includegraphics[width=5.1cm, height=3.5cm]{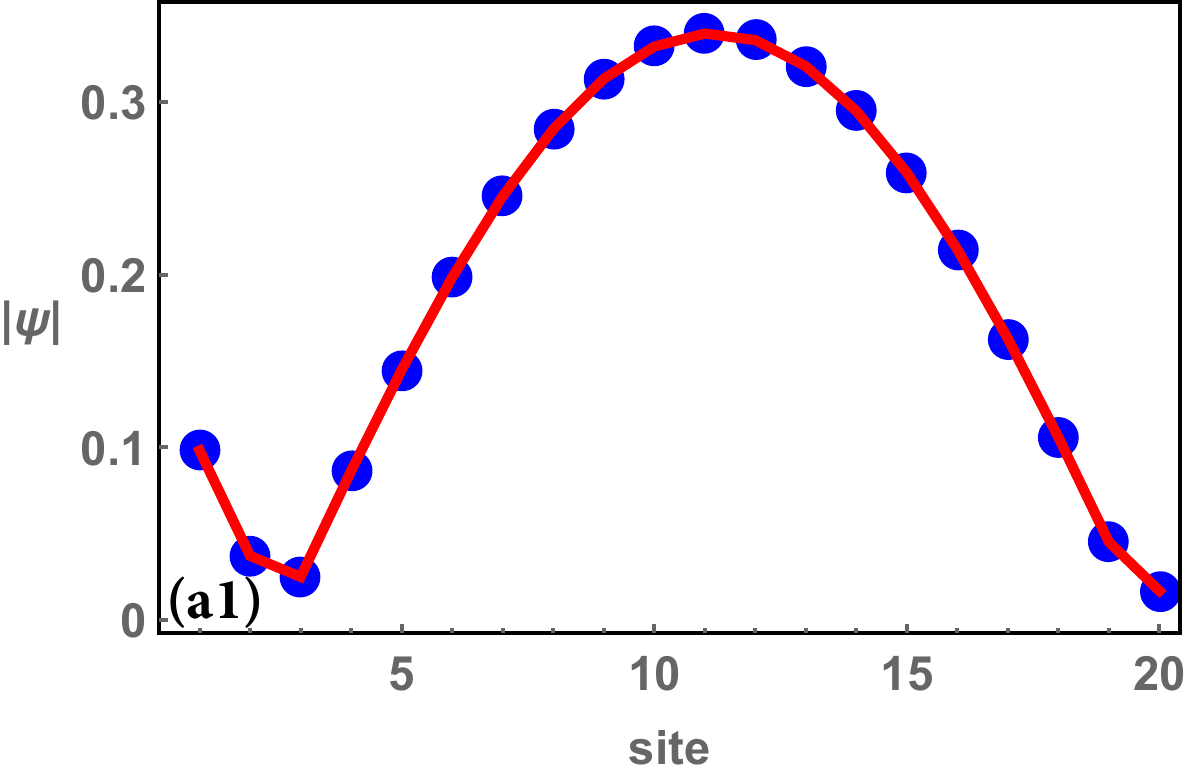}}
    \subfigure{\includegraphics[width=5cm, height=3.5cm]{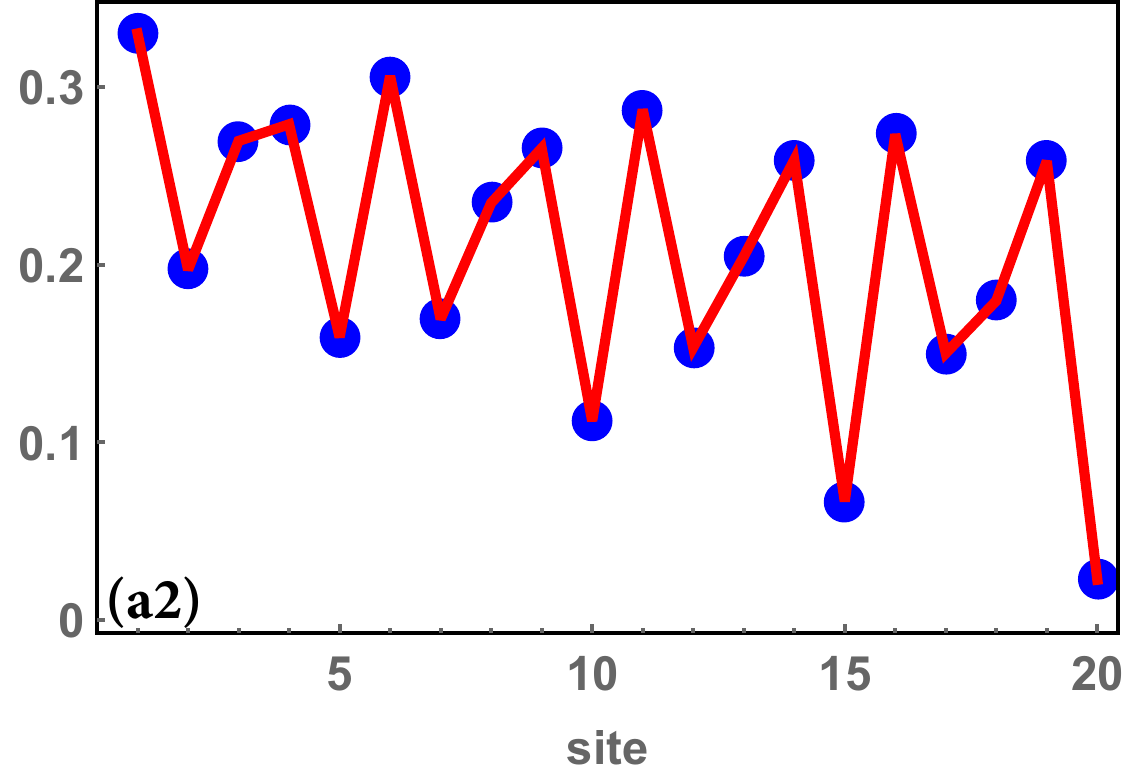}}
    \subfigure{\includegraphics[width=5cm, height=3.5cm]{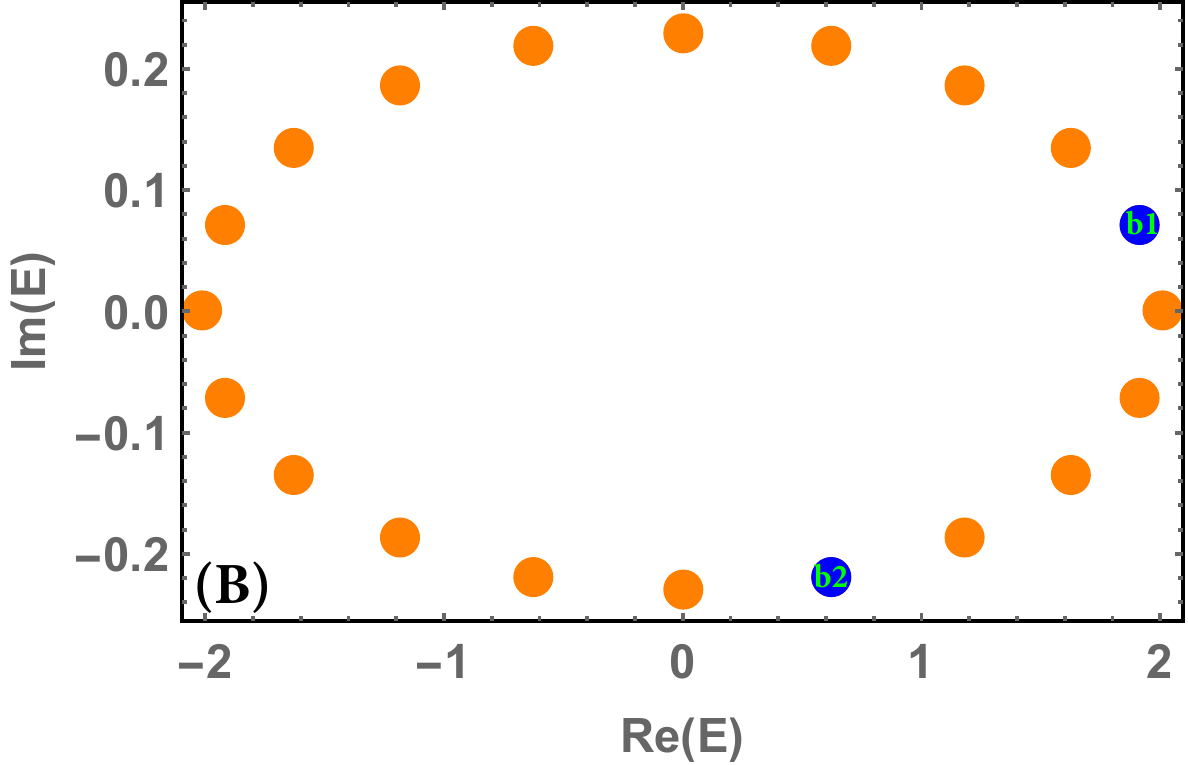}}
    \subfigure{\includegraphics[width=5.1cm, height=3.5cm]{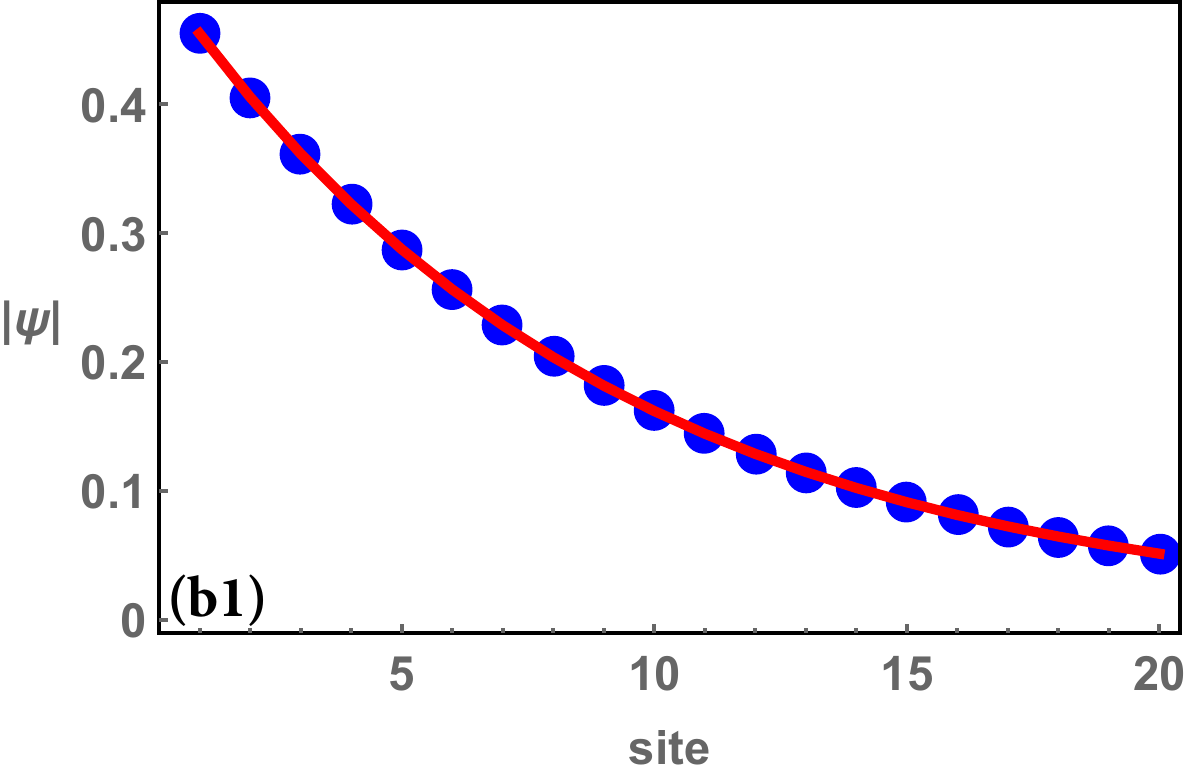}}
    \subfigure{\includegraphics[width=5cm, height=3.5cm]{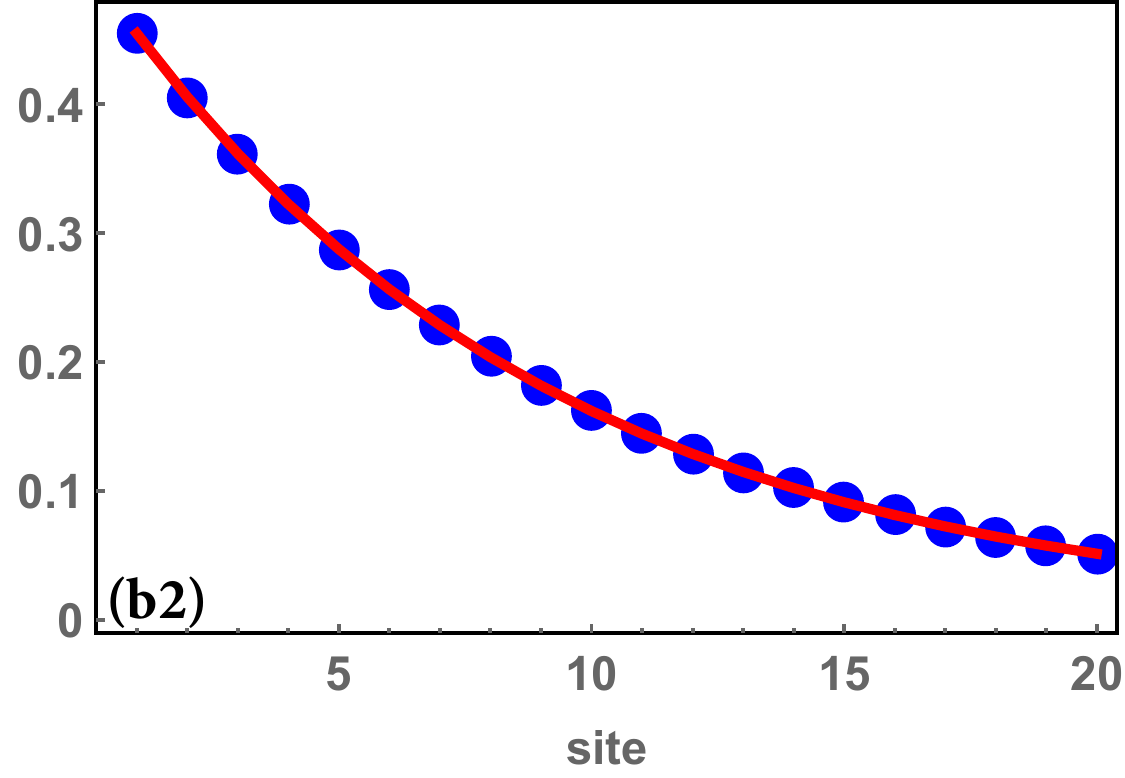}}
    \caption{The energy spectrum of Eq. (\ref{eqtargetham}) for (A) $\gamma=4.2$ and (B) $\gamma=\sqrt{24}$, respectively. $t=1$, $\delta=5$, and $N=20$. (a1) and (a2) The analytic (red lines) and numerical (blue dots) eigenstates corresponding to the blue points in (A). (b1) and (b2) The analytic (red lines) and numerical (blue dots) eigenstates corresponding to the blue points in (B). }
    \label{Fig-hnhssf}
\end{figure*}

\begin{figure} 
    \centering 
    \subfigure[]{\includegraphics[width=4.4cm, height=3cm]{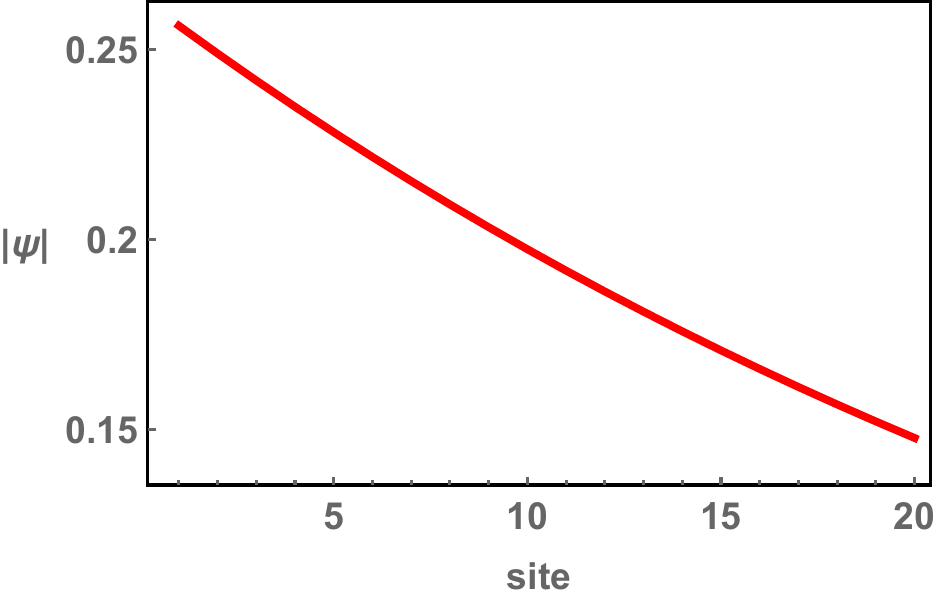}}
    \subfigure[]{\includegraphics[width=4.1cm, height=3.05cm]{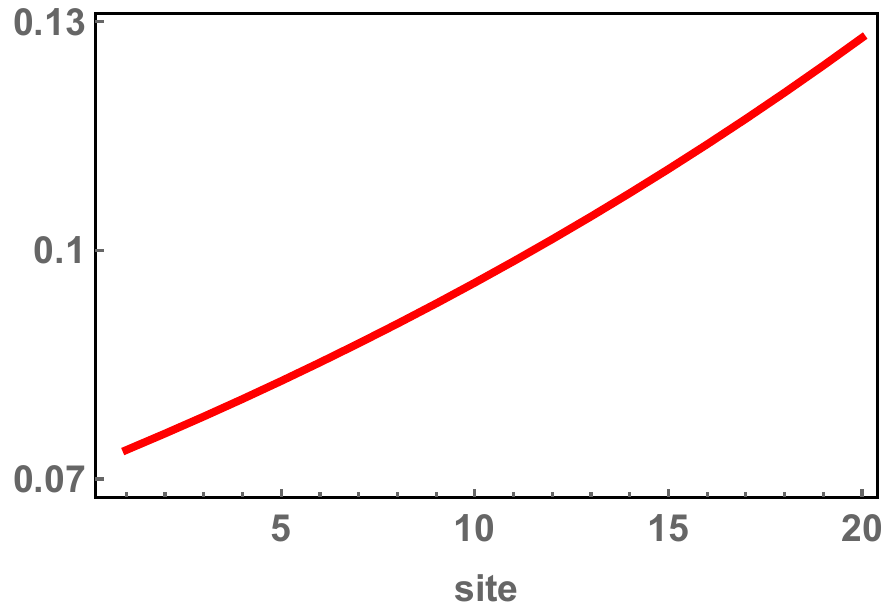}}
    \caption{(a) The LA component and (b) the RA component of the bidirectional SF mode in Fig. \ref{Fig-hnhssf}(a2). }
    \label{Fig-hnhssfcomp}
\end{figure}

\subsection{The emergent hybrid SFS effect with the boundary impurity}
\label{section3b}

Next, we consider the case with a nonsingular transfer matrix, whose energy is given by the expression
\begin{align}
    \label{eqhnhssfenformu}
    \varepsilon=2\sqrt{t_{L}t_{R}}\cos{\phi},\quad \phi=\phi_{R}+i\phi_{I}\in\mathbb{C},
\end{align}
following Eq. (\ref{eqzequalcos}). The solutions of $\phi$ lie in $\left[0,\pi\right]$ under OBC, leading to a pure skin effect in finite-size systems (see Appendix \ref{appendixd2}). To unambiguously establish the hybrid SFS effect, we perform a generalized gauge transformation on the Hamiltonian matrix $S^{-1}\mathcal{H}_{hn}S$ to cancel out the more dominant skin effect amplification factor $\Gamma^{n/2}$, resulting in \footnote{We have kept the notations of the annihilation and creation operators after the generalized gauge transformation. }
\begin{align}
    \label{eqtargetham}
    \hat{\mathcal{H}}_{it}=\sum_{n=1}^{N-1}t\left(\hat{c}_{n}^{\dagger}\hat{c}_{n+1}+\text{h.c.}\right)+(\delta+\gamma)\hat{c}_{1}^{\dagger}\hat{c}_{N}+(\delta-\gamma)\hat{c}_{N}^{\dagger}\hat{c}_{1},
\end{align}
where $S=\text{diag}\left\{r,r^{2},\ldots,r^{N}\right\}$ is the transformation matrix, $t=\sqrt{t_{L}t_{R}}$, $r=\sqrt{t_{R}/t_{L}}$, and we have set $r^{N-1}\gamma_{R}=\delta+\gamma$ and $r^{-(N-1)}\gamma_{L}=\delta-\gamma$ for fixed $t_{L}$ and $t_{R}$. The resulting Eq. (\ref{eqtargetham}) is exactly the Hamiltonian focused in Ref. \cite{guo2023scale}, a Hermitian Hamiltonian with a non-Hermitian boundary impurity $(\delta+\gamma)c_{1}^{\dagger}c_{N}+(\delta-\gamma)c_{N}^{\dagger}c_{1}$. Its transfer matrix becomes 
\begin{align}
    \label{eqapphnittransfermat}
    T_{it}=\left(\begin{matrix}
        \frac{\varepsilon}{t}&-1\\
        1&0
    \end{matrix}\right),
\end{align}
with $\Gamma_{it}=1$ and $\Delta_{it}=\varepsilon/t$, and the physical condition of $\mathcal{H}_{it}$ is the same as that of $\hat{\mathcal{H}}_{hn}$ with a nonsingular transfer matrix (see Appendix \ref{appendixd2})
\begin{align}
    \label{eqhnhssfcond4}
    &\left(\frac{\gamma_{L}}{t_{L}}\Gamma^{-\frac{N}{2}}+\frac{\gamma_{R}}{t_{R}}\Gamma^{\frac{N}{2}}\right)\sin{\phi}\nonumber\\
    &=\sin{\left((N+1)\phi\right)}-\frac{\gamma_{L}\gamma_{R}}{t_{L}t_{R}}\sin{\left((N-1)\phi\right)}.
\end{align}

In the PT-broken region $\gamma\in\left[|\delta-t|,\delta+t\right]$ with the emergence of complex energies, the solutions of the physical condition include coexisting real and complex values. Especially, at $\gamma=\gamma_{a}=\sqrt{\delta^{2}-t^{2}}$, the complex solutions are $\phi=2\pi m/N-i\log{\mu}/N$, with $m=1,2\ldots,N$ and $\mu=\delta/t+\sqrt{(\delta/t)^{2}-1}$ \cite{guo2023scale}. Subsequently, the eigenstate corresponding to energy $\varepsilon=2t\cos{\phi}$ of the Hamiltonian Eq. (\ref{eqtargetham}) reads 
\begin{align}
    \label{eqhnitvecs}
    \left(\begin{matrix}
        \psi_{n+1}\\\psi_{n}
    \end{matrix}\right)=\left[A_{L}(\phi)e^{i(n-1)\phi}+A_{R}(\phi)e^{-i(n-1)\phi}\right]\left(\begin{matrix}
        \psi_{2}\\\psi_{1}
    \end{matrix}\right), 
\end{align}
where the coefficients $A_{L,R}(\phi)$ are 
\begin{align}
    \label{eqhnitvecoff}
    A_{L}(\phi)&=\frac{T_{it}-e^{-i\phi}\mathbbm{1}}{2i\sin{\phi}},\nonumber\\
    A_{R}(\phi)&=\frac{-T_{it}+e^{i\phi}\mathbbm{1}}{2i\sin{\phi}}.
\end{align}

In Figs.~\ref{Fig-hnhssf}(A)(B), we illustrate the energy spectrum for two typical points in the PT-broken region $\gamma\in\left[|\delta-t|,\delta+t\right]$. The real (complex) energy (blue dots) in Fig.~\ref{Fig-hnhssf}(A) corresponds to the extended state (bidirectional SF mode), of which the perfectly matched analytic (red lines) and numerical (blue dots) results are plotted in Figs. \ref{Fig-hnhssf}(a1)(a2). Further, the SF distributions of the LA [$A_{R}(\phi)e^{-i(n-1)\phi}$] and RA [$A_{L}(\phi)e^{i(n-1)\phi}$] components of the bidirectional SF mode [Fig. \ref{Fig-hnhssf}(a2) with $\phi_{I}<0$] are illustrated in Fig. \ref{Fig-hnhssfcomp}. The two complex energies (blue dots) in Fig.~\ref{Fig-hnhssf}(B) correspond to the unidirectional SF modes (vanishing RA term), of which the perfectly matched analytic (red lines) and numerical (blue dots) results are plotted in Figs. \ref{Fig-hnhssf}(b1)(b2). Therefore, we observe the pure skin effect (the hybrid SFS effect) of $\hat{\mathcal{H}}_{hn}$ ($t_{L}\neq t_{R}$) corresponding to extended states (pure SF modes) with real (complex) $\phi$ solutions of $\hat{\mathcal{H}}_{it}$ after the inverse gauge transformation. We emphasize that the hybrid SFS effect with the boundary impurity may be a finite-size phenomenon since the thermodynamic limit may prevent the skin effect and SF effect simultaneously. The NHSE is usually sensitive to boundary conditions and vanishes in the thermodynamic limit with boundary impurities \cite{guo2021exact}. However, finite boundary impurities are essential for the emergence of the SF effect. Therefore, we can have a skin effect in the thermodynamic limit (with OBCs) or a finite-size SF effect (with finite boundary impurities) but not both simultaneously. 

\begin{figure*}
    \centering
    \subfigure{\includegraphics[width=6cm, height=5.7cm]{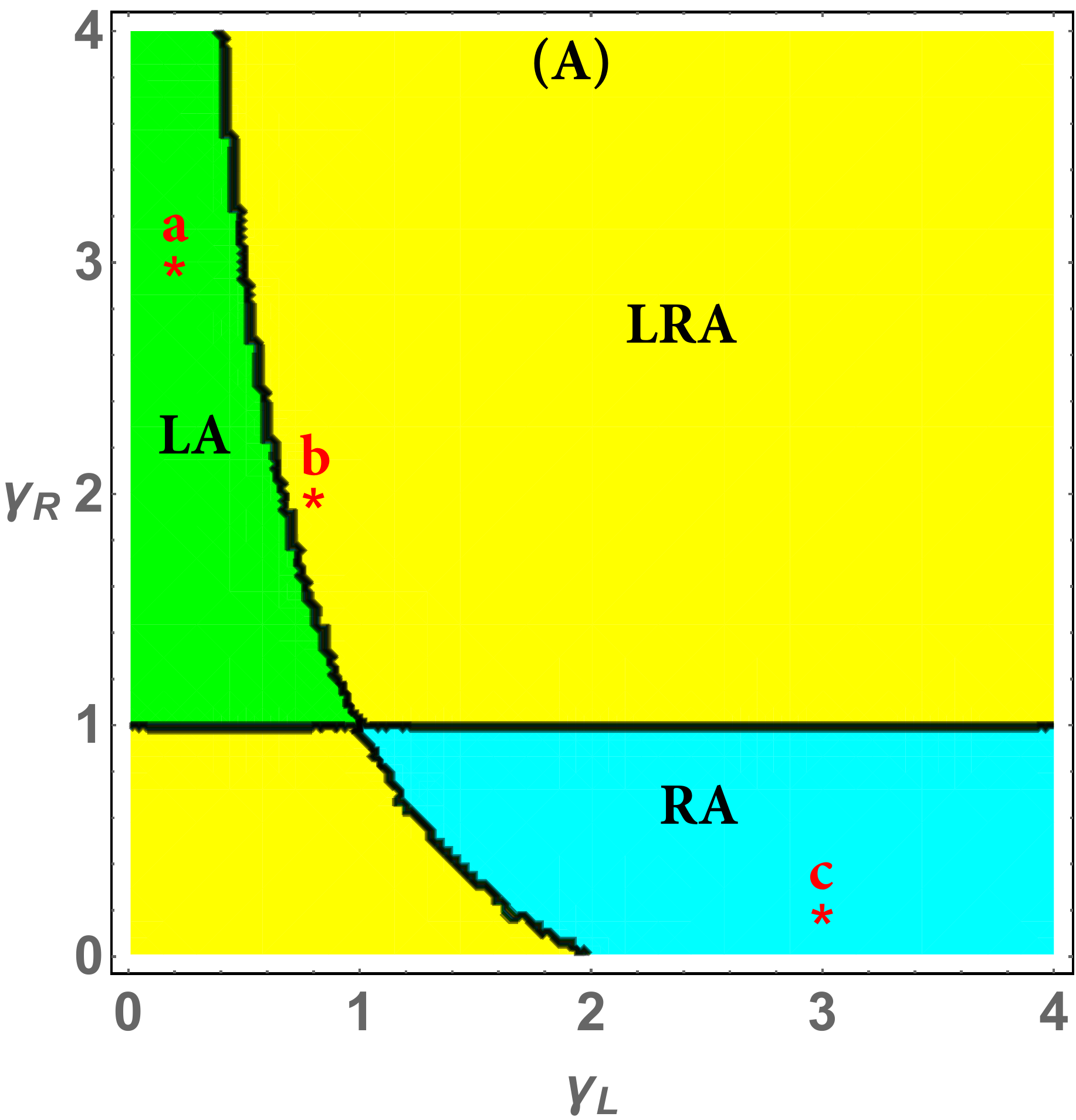}}
    \subfigure{\includegraphics[width=11.5cm, height=6cm]{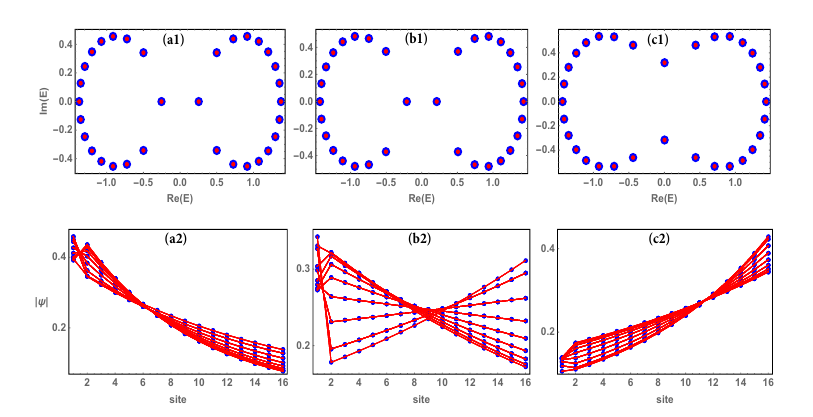}}
    \caption{(A) The phase diagram of the Hamiltonian $\hat{\mathcal{H}}_{nssh}$ with a singular transfer matrix, where the green, cyan, and yellow regions label the LA, RA, and LRA of the unidirectional SF modes, respectively. (a1)-(c1) The full analytic (red dots) and numerical (blue dots) energy spectra with $(\gamma_{L},\gamma_{R})=(0.2,3)$, $(0.8,2)$, and $(3,0.2)$, corresponding to red stars a, b, and c in (A), respectively. (a2)-(c2) The analytic (red lines) and numerical (blue dots) eigenstates corresponding to (a1)-(c1), respectively. $t_{1}=\gamma=0.5$, $t_{2}=1$, and $N=16$.}
    \label{Fig-nsshsing}
\end{figure*}

\begin{figure}
    \centering
    \subfigure[]{\includegraphics[width=4.2cm, height=3.2cm]{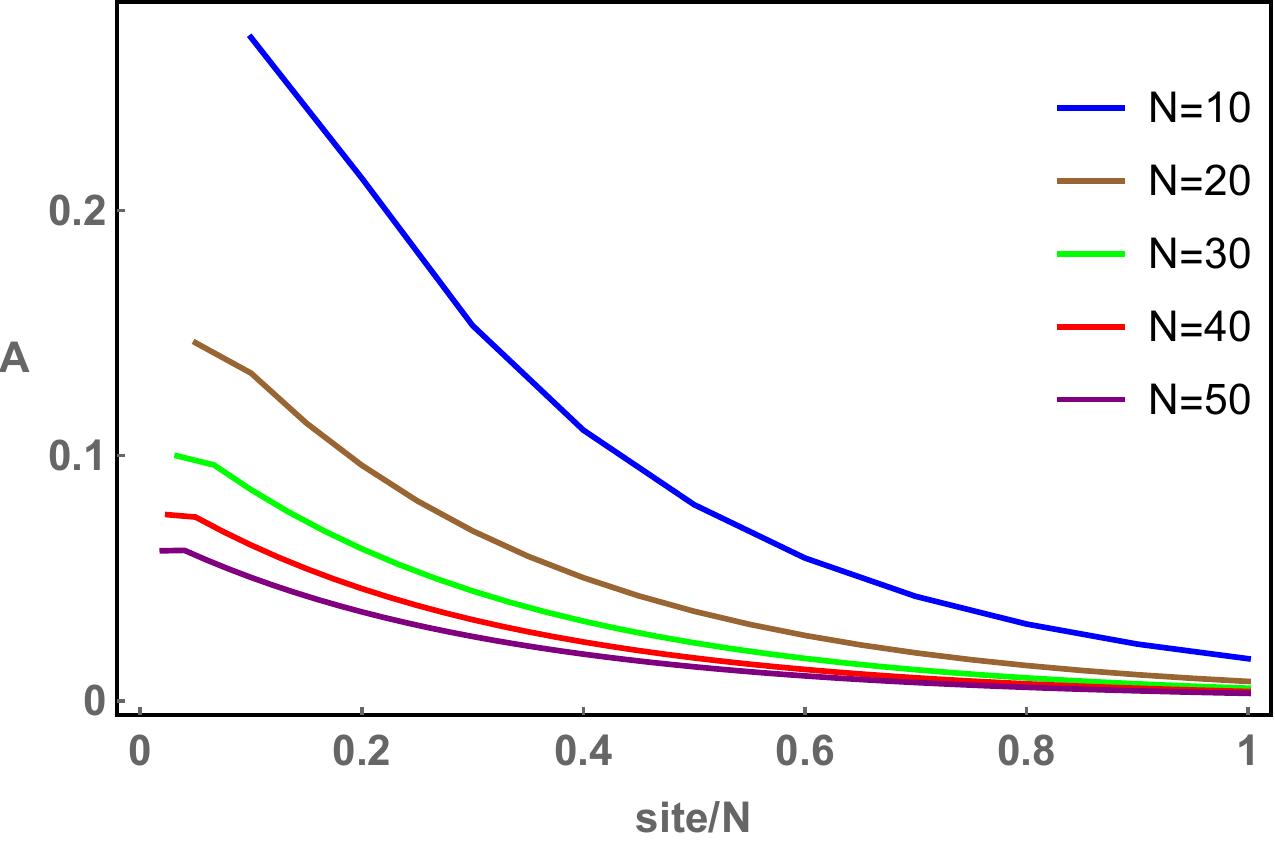}}
    \subfigure[]{\includegraphics[width=4cm, height=3.2cm]{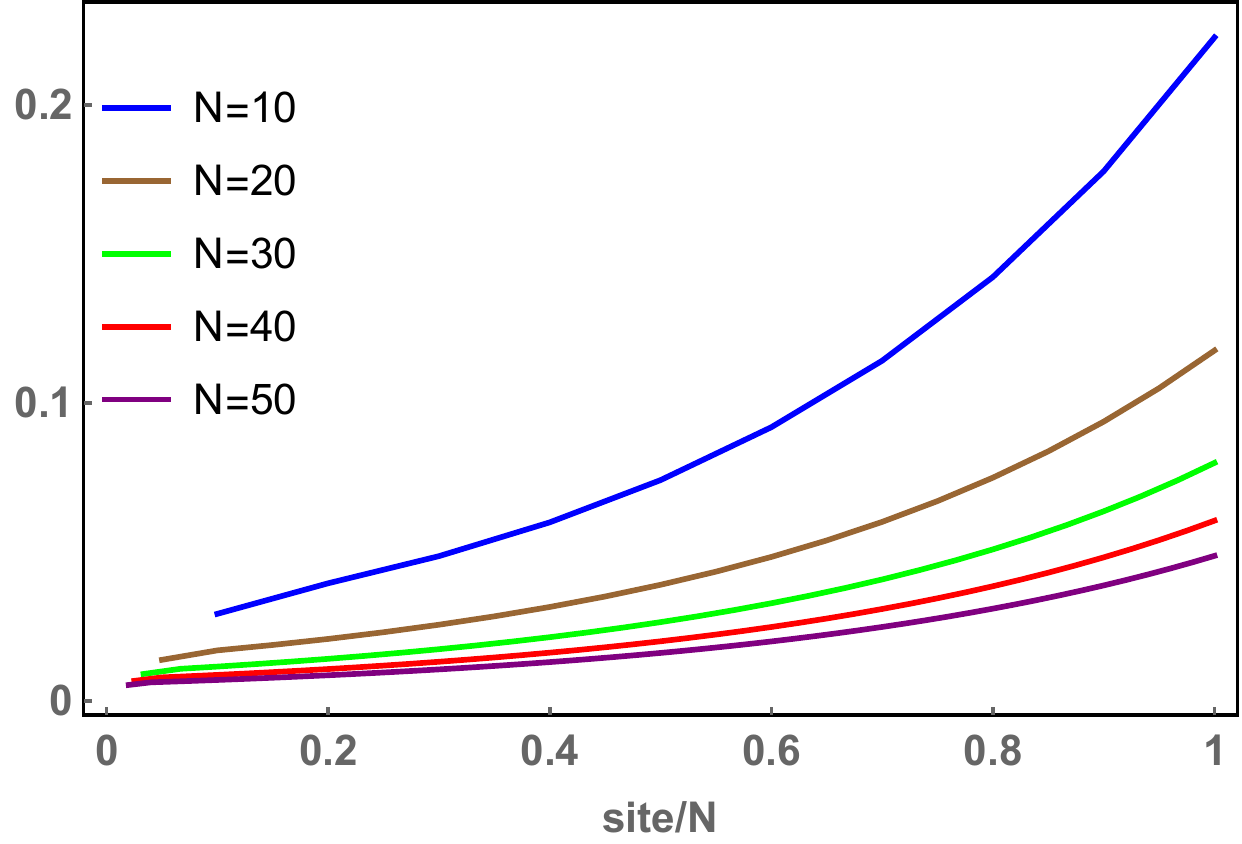}}
    \caption{The MASEs of (a) LA and (b) RA unidirectional SF modes for parameters corresponding to stars a and c in Fig. \ref{Fig-nsshsing}(A) and various system sizes $N$ display a quasilinearly dependent localization length. }
    \label{Fig-nsshsingscale}
\end{figure}

\section{The emergent unidirectional SF and hybrid SFS effects in the NH-SSH model}
\label{section4}

The well-known NH-SSH model is a typical $r=1$ Hamiltonian ($\hat{\mathcal{H}}_{nssh}$) from Eq. (\ref{eqsimplereduceham}) with on-site and hopping matrices 
\begin{align}
    \label{eqnsshham}
    \boldsymbol{M}=\left(\begin{matrix}
        0&t_{1}+\gamma\\
        t_{1}-\gamma&0
    \end{matrix}\right),\quad
    \boldsymbol{J}=\left(\begin{matrix}
        0&0\\
        t_{2}&0
    \end{matrix}\right),
\end{align}
where we assume $t_{1},t_{2},\gamma\geq 0$ for simplicity. The transfer matrix reads
\begin{align}
    \label{eqnsshtransfmat}
    T=\frac{1}{t_{2}(t_{1}+\gamma)}\left(\begin{matrix}
        \varepsilon^{2}-t_{1}^{2}+\gamma^{2}&-\varepsilon t_{2}\\
        \varepsilon t_{2}&-t_{2}^{2}
    \end{matrix}\right),
\end{align}
with
\begin{align}
    \label{eqnsshimptrace}
    \Delta&=\text{tr}\left(T\right)=\frac{\varepsilon^{2}-t_{1}^{2}+\gamma^{2}-t_{2}^{2}}{t_{2}(t_{1}+\gamma)},\nonumber\\
    \Gamma&=\det{\left(T\right)}=\frac{t_{1}-\gamma}{t_{1}+\gamma},
\end{align}
following the reduced SVD $\boldsymbol{J}=v\xi w^{\dagger}$, with $\xi=t_{2}$, $v=(0,1)^{T}$, and $w=(1,0)^{T}$. In the following, we introduce the boundary impurity in Eq. (\ref{eqboundimp}) to such an NH-SSH model and analyze its pure SF and hybrid SFS effects.

\subsection{The pure SF effect with boundary impurity}
\label{section4a}

The transfer matrix at the critical point $t_{1}=\gamma$ of the PT-symmetry transition (of the NH-SSH model under the OBC) is singular with $\Gamma=0$, where we can directly apply the analytic results in Sec. \ref{section2b1}. The single-particle eigenstates with respect to the two energy bands ($\pm$) are
\begin{align}
    \label{eqnsshsingvecs}
    \varepsilon_{m}^{\pm}&=\pm\sqrt{t_{2}^{2}+2t_{2}\gamma c_{m}^{-\frac{1}{N}}e^{-i\frac{2\pi m}{N}}},\nonumber\\
    \Psi_{n}^{\pm,m}&=\alpha_{n}^{\pm,m}v+\beta_{n}^{\pm,m}w,
\end{align}
where $m=1,2,\ldots,N$, 
\begin{align}
    \label{eqnsshsingsolu}
    \alpha_{n}^{\pm,m}&=c_{m}^{-\frac{n}{N}}e^{-i\frac{2\pi m}{N}n}\frac{c_{m}\gamma_{L} t_{2}}{\varepsilon_{m}^{\pm}},\quad n=1,2,\ldots,N-1,\nonumber\\
    \beta_{n}^{\pm,m}&=c_{m}^{-\frac{n-1}{N}}e^{-i\frac{2\pi m}{N}(n-1)}c_{m}\gamma_{L},\quad n=2,3,\ldots,N,\nonumber\\
    \alpha_{N}^{\pm,m}&=\frac{\varepsilon_{m}^{\pm\,2}-c_{m}\gamma_{L}(\varepsilon_{m}^{\pm\,2}-t_{2}^{2})}{\gamma_{R}\varepsilon_{m}^{\pm}t_{2}},\nonumber\\
    \beta_{1}^{\pm,m}&=1,
\end{align}
and $c_{m}$ satisfies the physical condition
\begin{align}
    \label{eqnsshsingcond}
    t_{2}(1-\gamma_{L}\gamma_{R})=2\gamma c_{m}^{-\frac{1}{N}}e^{-i\frac{2\pi m}{N}}(\gamma_{L}c_{m}-1).
\end{align}
These results indicate the emergence of the unidirectional SF effect with the localization length $\xi_{m}=N/|\text{Re}(\log{c_{m}})|$. We illustrate the resulting phase diagram in Fig.~\ref{Fig-nsshsing}(A). Applying Eq.~(\ref{eqnsshsingvecs}) to the three selected points (red stars) in Fig.~\ref{Fig-nsshsing}(A), which correspond to the unidirectional SF modes with LA, LRA, and RA, respectively, we obtain full energy spectra [red dots in Figs.~\ref{Fig-nsshsing}(a1)-(c1)] and eigenstates [red lines in Figs.~\ref{Fig-nsshsing}(a2)-(c2)] that perfectly match the numerical results (blue dots). As in Sec. \ref{section3a}, we illustrate the MASEs of the LA and RA unidirectional SF modes for various system sizes in Fig. \ref{Fig-nsshsingscale}, which suggests a quasilinear dependence where the localization length increases with the system size $N$. Again, we emphasize that the quasilinear dependence of the pure SF effect is fragile for large system sizes, which may gradually evolve into the pure skin effect in the thermodynamic limit (see Appendix \ref{appendixadd2} for details). 

\begin{figure*} 
    \centering 
    \subfigure[]{\includegraphics[width=5cm, height=3.5cm]{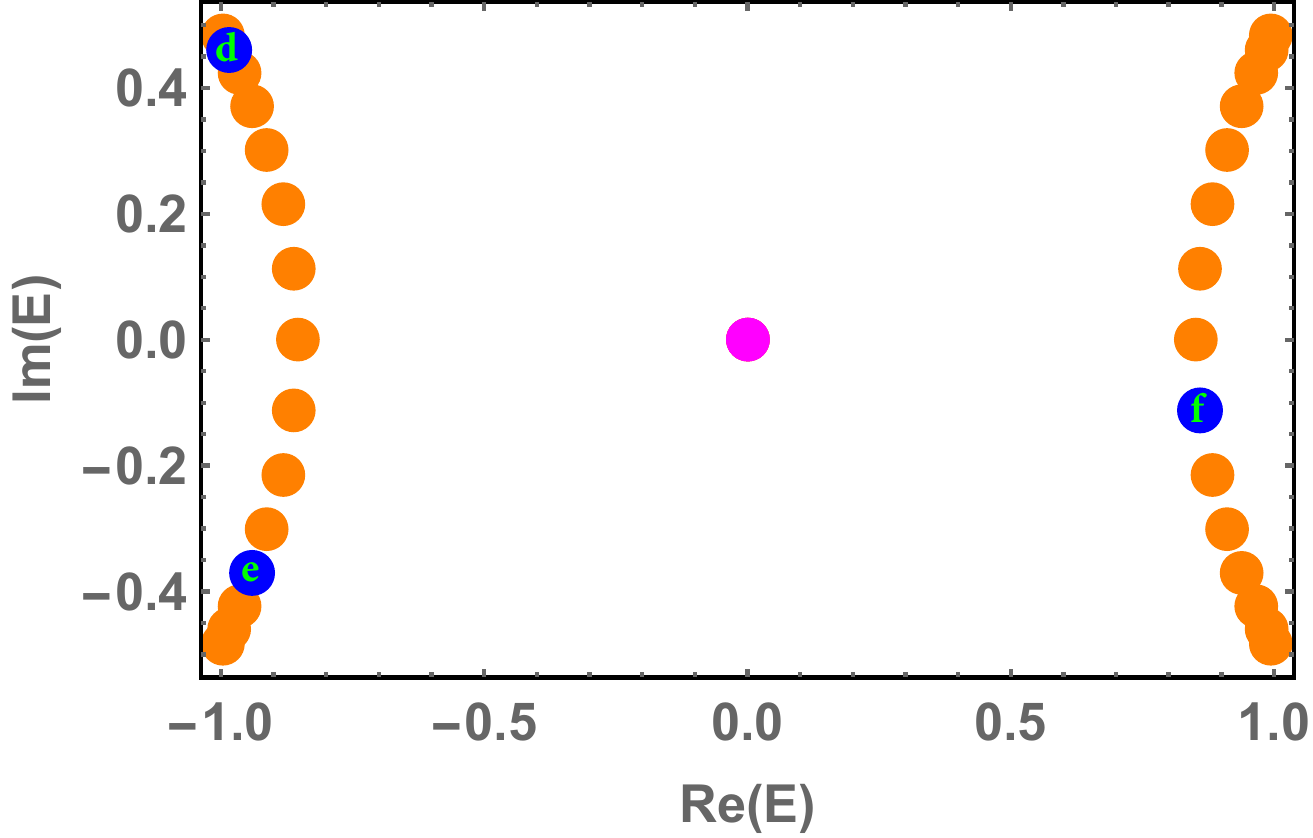}}
    \subfigure[]{\includegraphics[width=4.6cm, height=3.5cm]{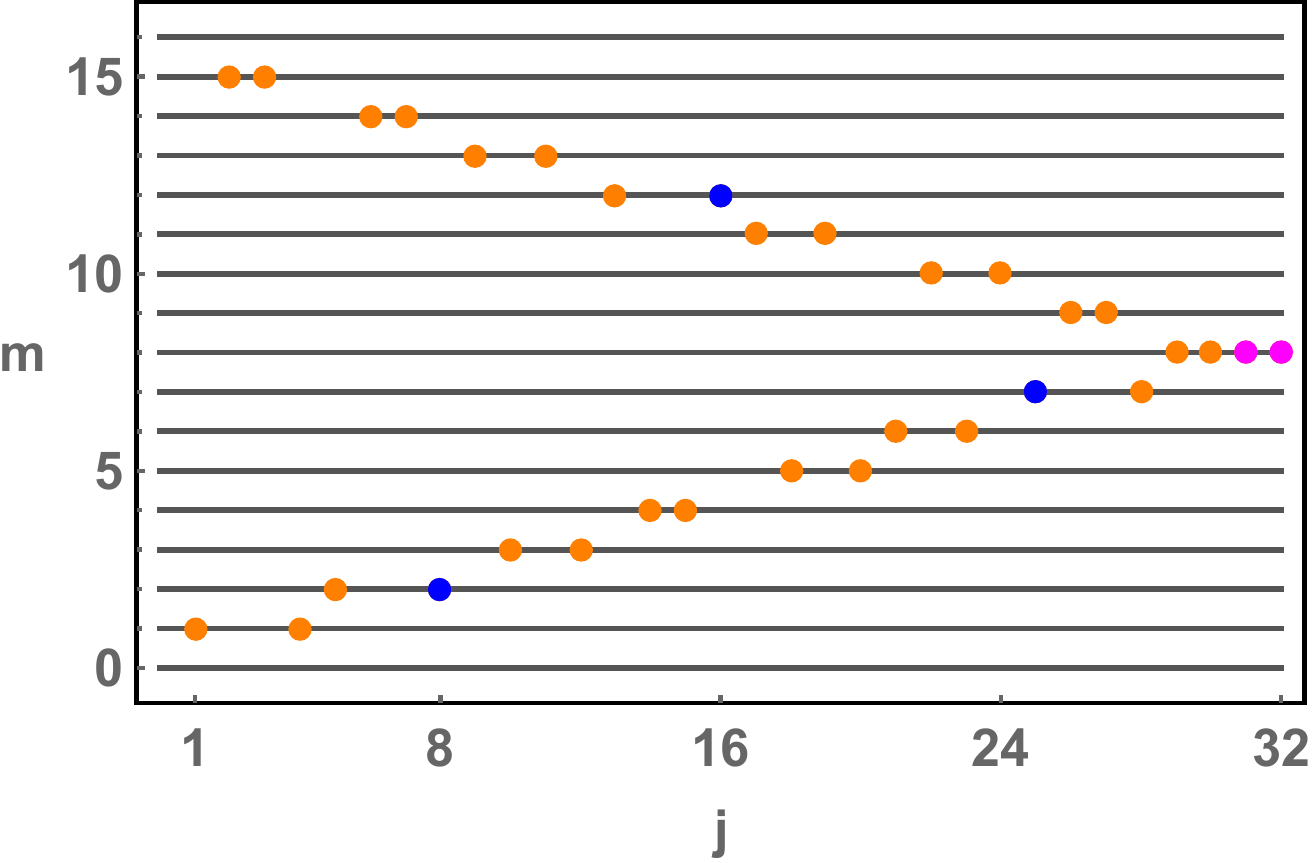}}
    \subfigure[]{\includegraphics[width=4.6cm, height=3.5cm]{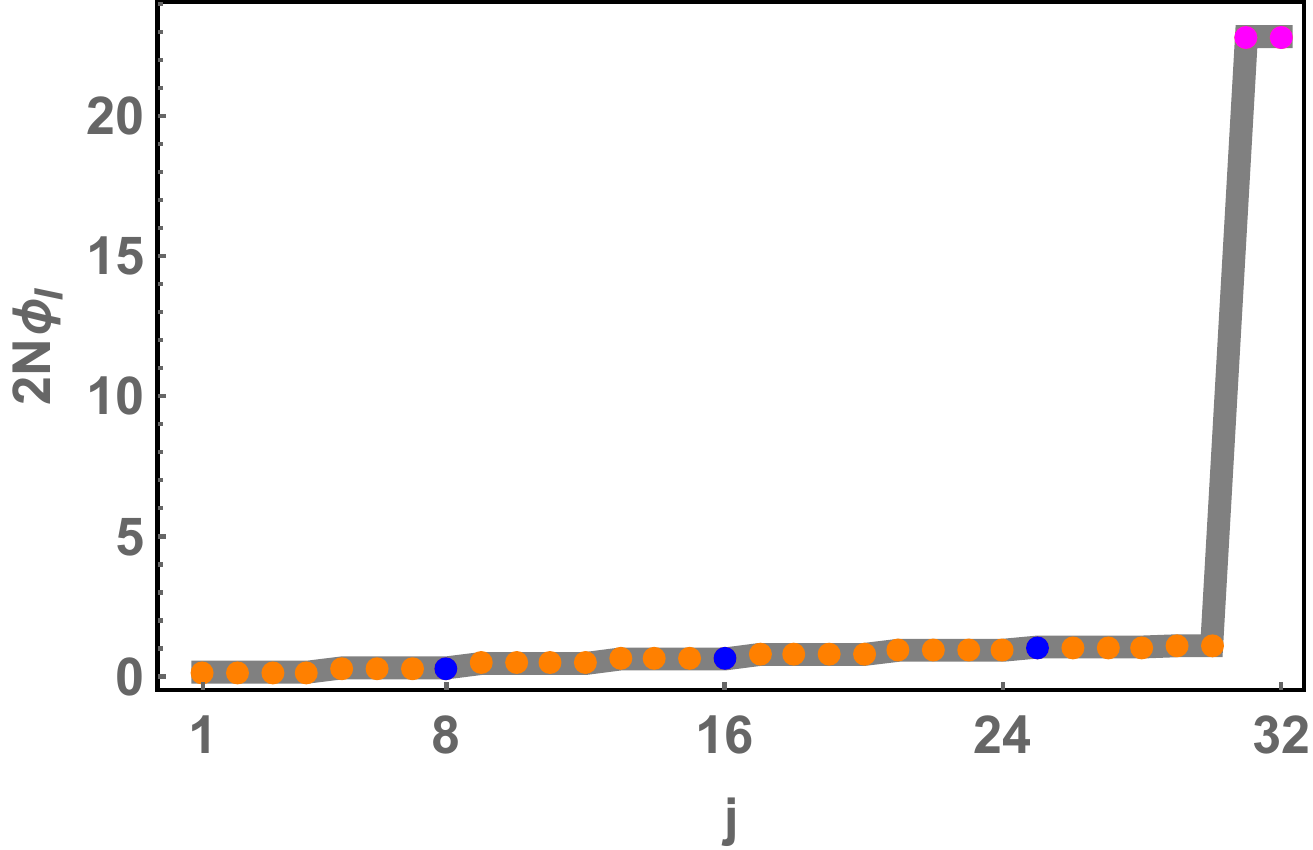}}
    \subfigure[]{\includegraphics[width=4.8cm, height=3.5cm]{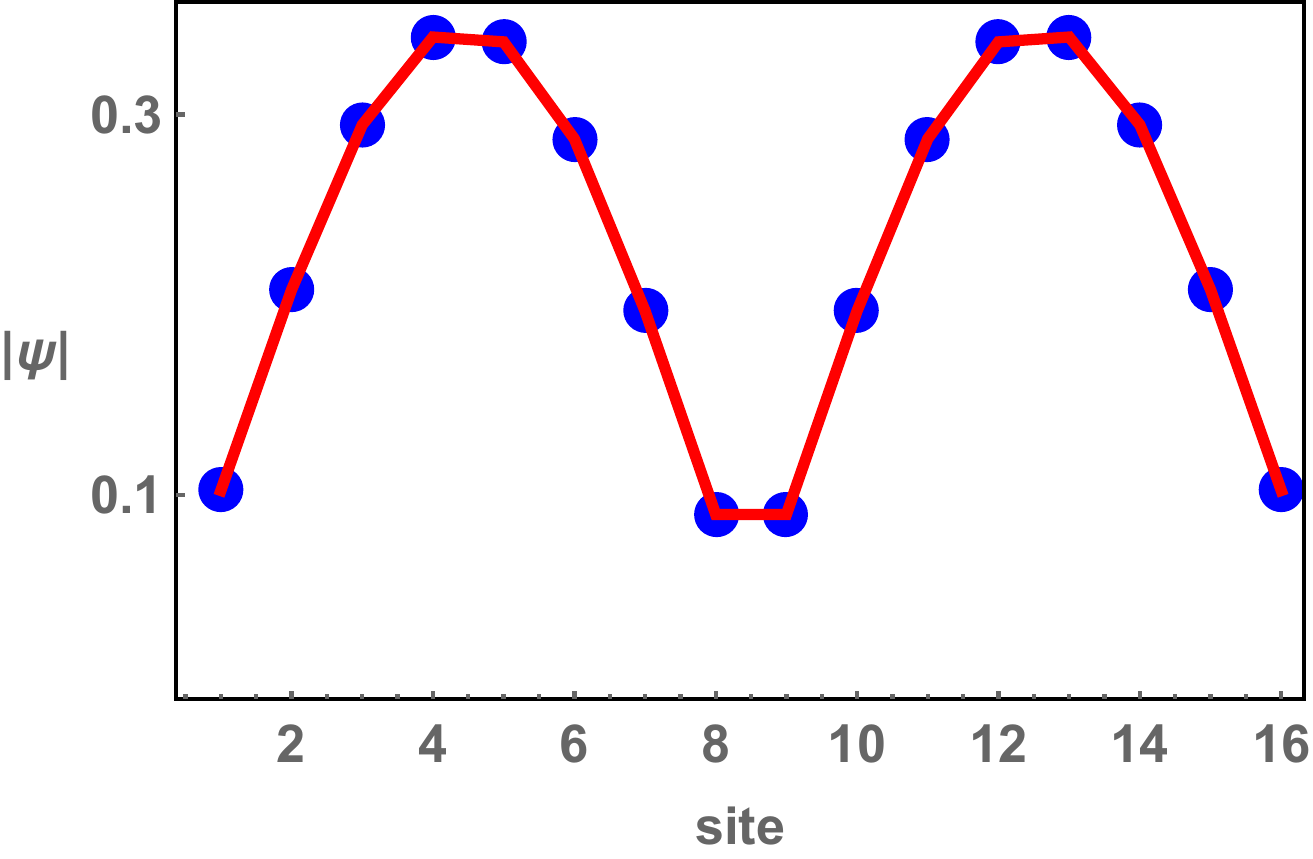}}
    \subfigure[]{\includegraphics[width=4.7cm, height=3.5cm]{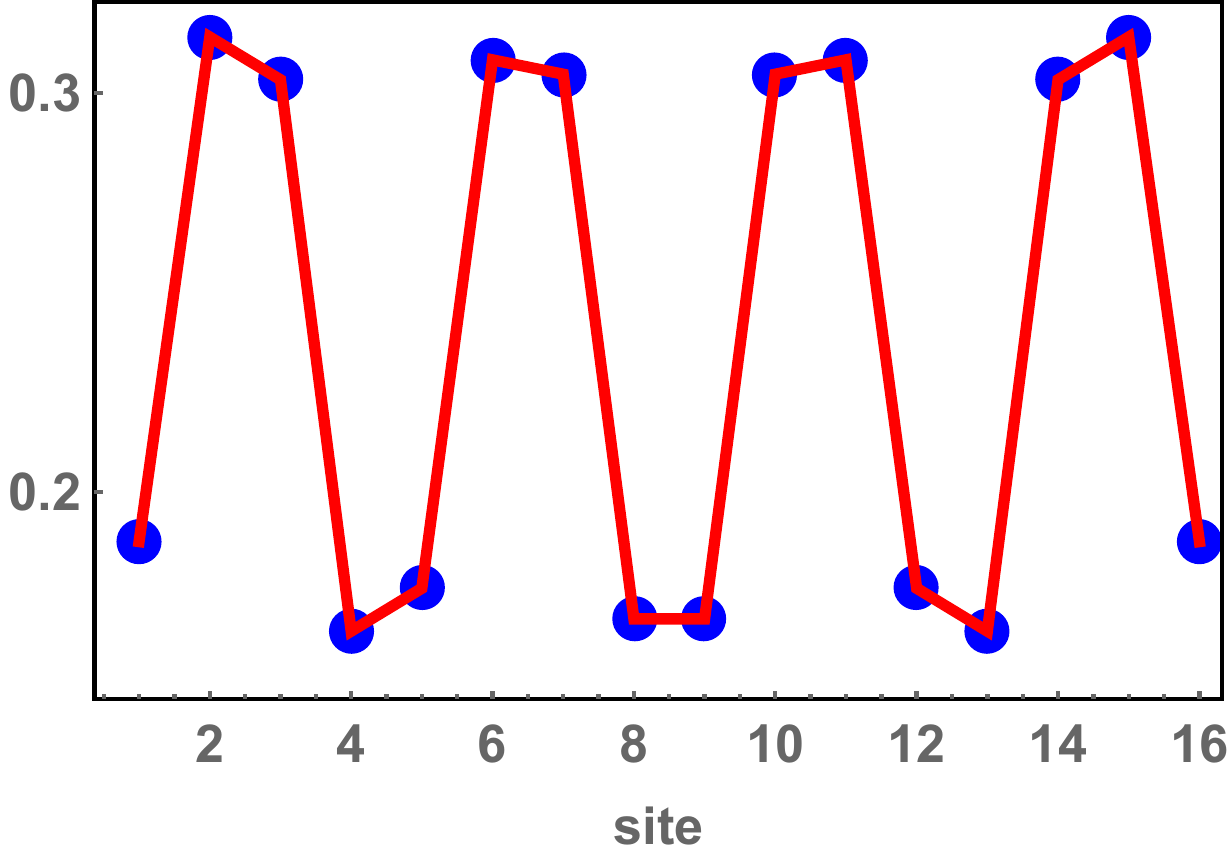}}
    \subfigure[]{\includegraphics[width=4.7cm, height=3.5cm]{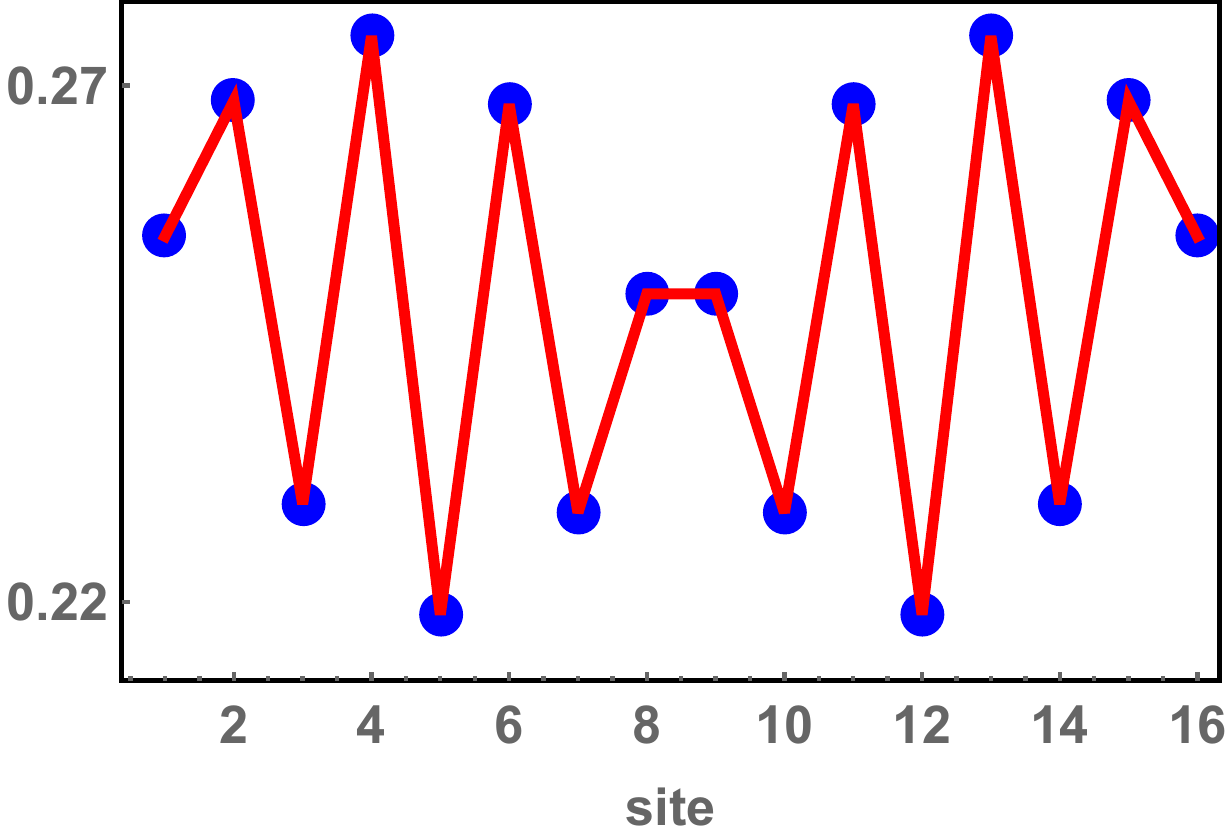}}
    \subfigure[]{\includegraphics[width=5cm, height=3.63cm]{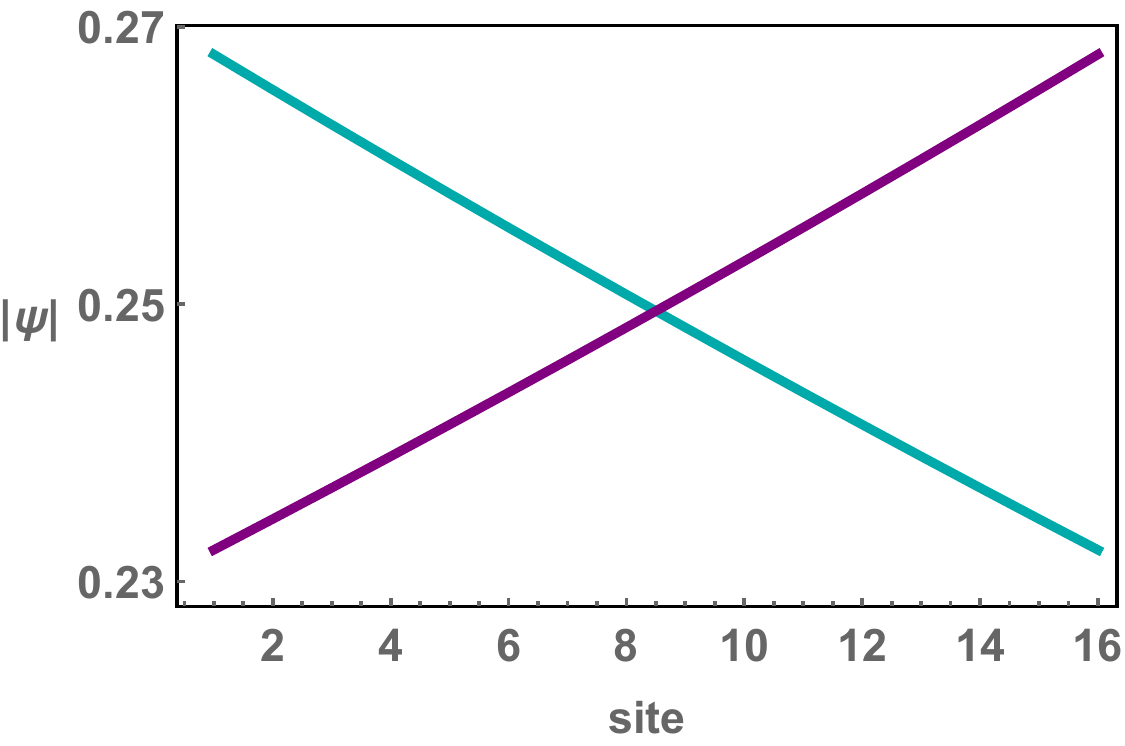}}
    \subfigure[]{\includegraphics[width=4.7cm, height=3.5cm]{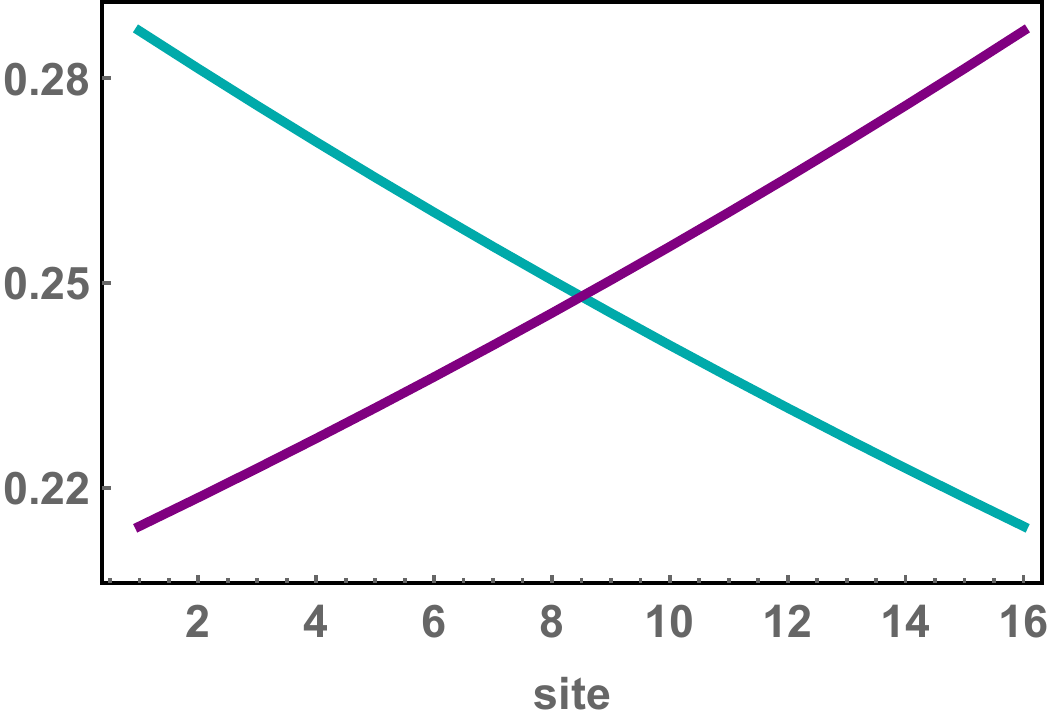}}
    \subfigure[]{\includegraphics[width=4.7cm, height=3.5cm]{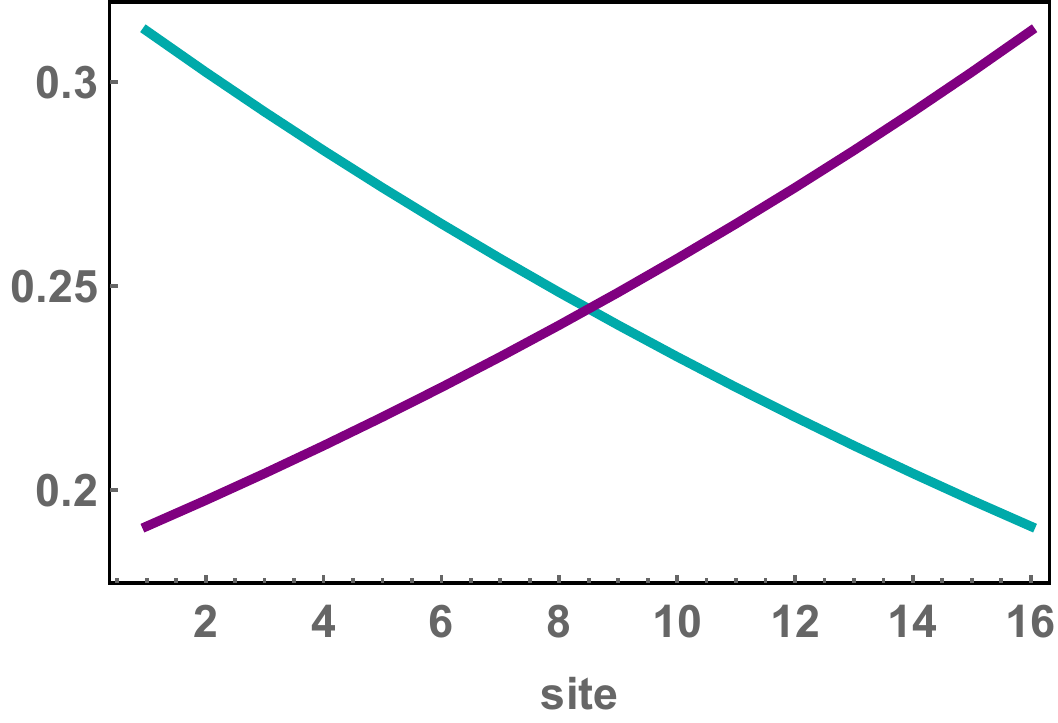}}
    \caption{(a) The energy spectrum of the NH-SSH model $\bar{\mathcal{H}}$ with the parameters $t_{1}=0.1$, $t_{2}=1$, $\gamma=0.5$, and $N=16$. 	 (b) All the solutions of $\phi_{R}$ in (a), labeled by $j$, fall on integer values $m=1,2,\ldots,16$ of $[2N\phi_{R}+\text{arg}(\mathscr{F})+2\pi p]/2\pi$, with $p=16$. (c) All the solutions of $2N\phi_{I}$ in (a) are also consistent with the analytic values $\log{|\mathscr{F}|}$. The magenta dots in (a)-(c) are related to the zero-energy modes. (d)-(f) The analytic (red lines) and numerical (blue dots) eigenstates corresponding to the blue dots in (a)-(c) match consistently. (g)-(i) The LA (cyan) and RA (purple) components of the bidirectional SF modes in (d)-(f), respectively, exhibit characteristic distributions. }
    \label{Fig-nsshhssf}
\end{figure*}

\subsection{The hybrid SFS effect under OBC}
\label{section4b}

The most counterintuitive phenomenon of the hybrid SFS effect emerges in the case of a nonsingular transfer matrix under the OBC. According to Appendix \ref{appendixb}, the physical condition is
\begin{align}
    \label{eqnsshhssfcond}
    \frac{\sin{\left(N\phi\right)}}{\sin{\left((N-1)\phi\right)}}=\frac{t_{2}\sqrt{(t_{1}+\gamma)(t_{1}-\gamma)}}{\varepsilon^{2}-t_{1}^{2}+\gamma^{2}}\equiv q,
\end{align}
which leads to the solutions of $\phi=\phi_{R}+i\phi_{I}$,
\begin{align}
    \phi_{I}&=\frac{\log{|\mathscr{F}|}}{2N},\label{eqnsshhsshphiim}\\
    \phi_{R}&=\frac{2\pi m-\text{arg}(\mathscr{F})}{2N},\label{eqnsshhsshphire}\\
    \mathscr{F}&=\frac{1-qe^{-i\phi}}{1-qe^{i\phi}},\label{eqnsshhssfphisolu}
\end{align}
with $m=1,2,\ldots,N$. In the region $t_{1}>\gamma$, the real $\phi$ and OBC energy spectrum exist simultaneously. However, in the region $t_{1}<\gamma$, complex energies lead to the presence of an imaginary part of $\phi$, namely, the quasi-$1/N$ proportion (\ref{eqnsshhsshphiim}). Hence, the hybrid SFS effect, newly established in this work, emerges following Eq. (\ref{eqobceigenstate}). 

As in Sec. \ref{section3b}, we perform a generalized gauge transformation $S=\text{diag}\left\{1,r,r,\ldots,r^{N-1},r^{N-1},r^{N}\right\}$, with $r=\sqrt{\Gamma}$ for $\hat{\mathcal{H}}_{nssh}$, to cancel out the amplification factor of the NHSE, which results in a Hamiltonian $\bar{\mathcal{H}}$ in the form of Eq. (\ref{eqsimplereduceham}) with the on-site and hopping matrices
\begin{align}
    \label{eqnsshhambar}
    \bar{M}=\left(\begin{matrix}
        0&\bar{t}_{1}\\
        \bar{t}_{1}&0
    \end{matrix}\right),\quad
    \bar{J}=\boldsymbol{J}, \quad \bar{t}_{1}=\sqrt{(t_{1}+\gamma)(t_{1}-\gamma)}.
\end{align}

Subsequently, we obtain $\bar{\Delta}=(\varepsilon^{2}-\bar{t}_{1}^{2}-t_{2}^{2})/\bar{t}_{1}t_{2}$ and $\bar{\Gamma}=1$. The physical condition remains the same as in Eq.~(\ref{eqnsshhssfcond}) for $\hat{\mathcal{H}}_{nssh}$. Together with Eq. (\ref{eqzequalcos}), we obtain the energy 
\begin{align}
    \label{eqnsshhssfenexp}
    \varepsilon^{2}=t_{1}^{2}-\gamma^{2}+t_{2}^{2}+2t_{2}\bar{t}_{1}\cos{\phi},
\end{align}
whose corresponding eigenstates read 
\begin{align}
    \label{eqnsshbarvecs}
    \bar{\Psi}_{n}=\bar{\mathscr{A}}_{L}(\phi)e^{in\phi_{R}}e^{-n\phi_{I}}+\bar{\mathscr{A}}_{R}(\phi)e^{-in\phi_{R}}e^{n\phi_{I}},
\end{align}
where the coefficients are given by
\begin{align}
    \label{eqnsshbarcoof}
    \bar{\mathscr{A}}_{L}(\phi)&=\frac{1}{2iq\sin{\phi}}\left[t_{2}\bar{\mathcal{G}}_{vv}v+(e^{-i\phi}-qe^{-2i\phi})w\right],\nonumber\\
    \bar{\mathscr{A}}_{R}(\phi)&=\frac{1}{2iq\sin{\phi}}\left[-t_{2}\bar{\mathcal{G}}_{vv}v+(-e^{i\phi}+qe^{2i\phi})w\right],
\end{align}
with $\bar{\mathcal{G}}_{vv}=\varepsilon/(\varepsilon^{2}-t_{1}^{2}+\gamma^{2})$. 

Utilizing the energy spectrum in Fig.~\ref{Fig-nsshhssf}(a), we observe that the resulting solutions of $\phi$ and the corresponding eigenstates are perfectly consistent with the analytic expressions, i.e., Eqs.~(\ref{eqnsshhsshphiim})-(\ref{eqnsshhssfphisolu}), (\ref{eqnsshbarvecs}), and (\ref{eqnsshbarcoof}), as in Figs.~\ref{Fig-nsshhssf}(b)-(f). Also, in Figs. \ref{Fig-nsshhssf}(g)-(i), we illustrate the SF distributions of the LA [$\bar{\mathscr{A}}_{L}(\phi)e^{in\phi_{R}}e^{-n\phi_{I}}$] and RA [$\bar{\mathscr{A}}_{R}(\phi)e^{-in\phi_{R}}e^{n\phi_{I}}$] components of the bidirectional SF modes $\bar{\Psi}_{n}$ in Figs. \ref{Fig-nsshhssf}(d)-(f), which appear tp be extended and indicate the hybrid SFS effect of the original NH-SSH model before the generalized gauge transformation. 

Finally, as we turn on the boundary impurity, the physical condition becomes
\begin{align}
    \label{eqnsshhssfimpcond}
    &\left(\gamma_{L}\Gamma^{-\frac{N}{2}}+\gamma_{R}\Gamma^{\frac{N}{2}}\right)\sin{\phi}=\sin{\left((N+1)\phi\right)}\nonumber\\
    &+\frac{t_{2}(1-\gamma_{L}\gamma_{R})}{\sqrt{(t_{1}+\gamma)(t_{1}-\gamma)}}\sin{(N\phi)}-\gamma_{L}\gamma_{R}\sin{\left((N-1)\phi\right)},
\end{align}
according to Eq. (\ref{eqnsshhssfenexp}) and Appendix \ref{appendixc}, and is equivalent to the result in Ref. \cite{guo2021exact}. With the available complex solutions of $\phi$, eigenstates following Eq. (\ref{eqimpbulkeigenstate}) emerge, realizing the hybrid SFS effect and being fully consistent with the numerical results (see Appendix \ref{appendixe}). Nevertheless, such a hybrid SFS effect with the boundary impurity is once again limited to finite-size systems.

\section{Conclusions and discussion}
\label{section5}

We have developed a unified theory of the skin effect and the SF effect in non-Hermitian systems with or without the boundary impurity from a transfer matrix perspective. We derived the analytic expressions for the single-particle bulk eigenstates and energy spectrum for 1D non-Hermitian systems with $2\times 2$ transfer matrices, which exhibit excellent consistency with the numerical results of the HN model and the NH-SSH model. The unidirectional SF effect accompanies a singular transfer matrix, while the hybrid SFS effect emerges with a nonsingular transfer matrix. The unidirectional SF effect portrays a localization length of the eigenstates that is quasilinear with respect to the system size, while the hybrid SFS effect shows coexistence of the skin effect and the SF effect. We further revealed that although the hybrid SFS effects may emerge in finite-size systems under the OBC, the skin effect eventually prevails over the SF effect as the system size increases, i.e., $\phi_{I}=0$ in the thermodynamic limit. We may then generate the energy spectrum from $\Delta=2\sqrt{\Gamma}\cos{\phi_{R}}$ [Eq. (\ref{eqzequalcos})]. Interestingly, the norm $\sqrt{|\Gamma|}$ with the corresponding $(\text{arg}(\Gamma),\phi_{R})$ outlines the GBZ and reproduces the non-Bloch band theory, opening a viable route for further studies and generalizations of non-Hermitian systems via the transfer matrix approach perspective. 

The constraints on the assumption $\boldsymbol{J}_{R}=\boldsymbol{J}_{L}$, rank-$1$ of $\boldsymbol{J}$ with $\mathcal{N}=2$, namely, $2\times 2$ transfer matrices, and the boundary impurity in Eq. (\ref{eqboundimp}) are sufficiently general to capture the core mechanism and behaviors of the pure SF and hybrid SFS effects here. Overall, our exploration of the pure SF and hybrid SFS effects complements the finite-size manifestation of the single-particle eigenstates and compensates for the localization behaviors of the skin effect in non-Hermitian systems. Nevertheless, many unanswered questions concerning the skin and SF effects exist, such as the role of and interplay between impurity and disorder in the bulk \cite{molignini2023anomalous,longhi2021disorder,okuma2021disorder,luo2021transfer,yuce202disorder,sarkar2022disorder}, the role of various symmetries \cite{gong2018,kawabataprx} and topological characters \cite{transfer3,transfer4,vatsal2016transfer,kunst2019transfer}, etc. As a straightforward extension of our conclusions to higher dimensions, we consider a two-dimensional non-Hermitian system with boundary impurities or OBCs that fully decouples along its separate directions. The pure SF effects (hybrid SFS effects) along both directions naturally introduce eigenstate decays that multiply and thus an SF effect (a hybrid SFS effect accompanying the higher-order skin effect) in the entire two-dimensional plane. Therefore, our work may extend the scope and revise the behaviors of the higher-order skin effect \cite{kawabata2019second,kawabatahigher,okugawa2020}, the hybrid skin-topological effect \cite{lee2019ho,fu2021}, and the geometry-dependent skin effect \cite{zhang2022uni,hu2023nonhermitian}, which are potentially observable in various non-Hermitian platforms, e.g., photonic devices \cite{optical3,optical4,optical5,lin2022mobilityedge,optical6}. More broadly, the pure SF and hybrid SFS effects may also influence non-Hermitian many-body physics \cite{manybody4,manybody10,manybody12}, such as ground-state corrections, the entanglement spectrum (entropy), critical exponents and scaling, and other topics in condensed matter physics.

\section*{Acknowledgements}
We acknowledge support from the National Key R\&D Program of China (No.2022YFA1403700) and the National Natural Science Foundation of China (No.12174008 \& No.92270102). 

\appendix

\begin{widetext}
\section{The boundary equations with the boundary impurity}
\label{appendixa}
Modifying Eq.~(\ref{eqrecursionrelation}) at the boundaries with the setting $\Psi_{0}\equiv\Psi_{N}$, we obtain
\begin{align}
    \label{eqappbounds}
    \Psi_{N}&=\mathcal{G}\boldsymbol{J}^{\dagger}\Psi_{N-1}+\gamma_{L}\mathcal{G}\boldsymbol{J}\Psi_{1}\nonumber,\\
    \Psi_{1}&=\gamma_{R}\mathcal{G}\boldsymbol{J}^{\dagger}\Psi_{N}+\mathcal{G}\boldsymbol{J}\Psi_{2}.
\end{align}
Combining the results with Eqs. (\ref{eqsvd})-(\ref{eqonsitgreenexp}), we obtain 
\begin{align}
    \label{eqappboundsdetail}
    \alpha_{N}&=\mathcal{G}_{wv}\xi\alpha_{N-1}+\gamma_{L}\mathcal{G}_{vv}\xi\beta_{1},\nonumber\\
    \beta_{N}&=\mathcal{G}_{ww}\xi\alpha_{N-1}+\gamma_{L}\mathcal{G}_{vw}\xi\beta_{1},\nonumber\\
    \alpha_{1}&=\gamma_{R}\mathcal{G}_{wv}\xi\alpha_{N}+\mathcal{G}_{vv}\xi\beta_{2},\nonumber\\
    \beta_{1}&=\gamma_{R}\mathcal{G}_{ww}\xi\alpha_{N}+\mathcal{G}_{vw}\xi\beta_{2}.
\end{align}   
Further derivation gives
\begin{align}
    \label{eqappboundsfinal}
    \left(\begin{matrix}
        \beta_{1}\\
        \alpha_{N}
    \end{matrix}
        \right)
    &=\left(\begin{matrix}
        \frac{1}{\gamma_{L}}\xi^{-1}\mathcal{G}_{vw}^{-1}&-\frac{1}{\gamma_{L}}\xi^{-1}\mathcal{G}_{vw}^{-1}\mathcal{G}_{ww}\xi\\
        \mathcal{G}_{vv}\mathcal{G}_{vw}^{-1}&\left(\mathcal{G}_{wv}-\mathcal{G}_{vv}\mathcal{G}_{vw}^{-1}\mathcal{G}_{ww}\right)\xi
    \end{matrix}
    \right)
    \left(\begin{matrix}
        \beta_{N}\\
        \alpha_{N-1}
    \end{matrix}
        \right),\nonumber\\
    \left(\begin{matrix}
        \beta_{2}\\
        \alpha_{1}
    \end{matrix}
        \right)
    &=\left(\begin{matrix}
        \xi^{-1}\mathcal{G}_{vw}^{-1}&-\gamma_{R}\xi^{-1}\mathcal{G}_{vw}^{-1}\mathcal{G}_{ww}\xi\\
        \mathcal{G}_{vv}\mathcal{G}_{vw}^{-1}&\gamma_{R}\left(\mathcal{G}_{wv}-\mathcal{G}_{vv}\mathcal{G}_{vw}^{-1}\mathcal{G}_{ww}\right)\xi
    \end{matrix}\right)
    \left(\begin{matrix}
        \beta_{1}\\
        \alpha_{N}
    \end{matrix}
        \right),
\end{align}
which are exactly the equations on the boundaries
\begin{align}
    \label{eqappboundsresult}
    \Phi_{1}&=\left(\begin{matrix}
        \frac{1}{\gamma_{L}}&0\\
        0&1
    \end{matrix}\right)
    T\Phi_{N}=K_{L}T\Phi_{N},\nonumber\\
    \Phi_{2}&=T\left(\begin{matrix}
        1&0\\
        0&\gamma_{R}
    \end{matrix}\right)\Phi_{1}=TK_{R}\Phi_{1}.
\end{align}

\section{The solutions of nonsingular cases under the OBC}
\label{appendixb}

The OBC refers to the hard or Dirichlet boundary condition $\Psi_{0}=\Psi_{N+1}=0$, implying $\alpha_{0}=\beta_{N+1}=0$, such that
\begin{align}
    \label{eqappobccond}
    \Phi_{1}=\left(\begin{matrix}\beta_{1}\\0
    \end{matrix}\right),\quad
    \Phi_{N+1}=\left(\begin{matrix}0\\\alpha_{N}
    \end{matrix}\right).
\end{align}
After normalization, we obtain the OBC equation
\begin{align}
    \label{eqappobceq}
    T^{N}\left(\begin{matrix}1\\0
    \end{matrix}\right)=\left(\begin{matrix}0\\\tau
    \end{matrix}\right),
\end{align}
with $\tau\in\mathbb{C}$. Accordingly, the propagating relation of $\Phi_{n}$ is illustrated in Fig. \ref{Fig-propageting}(b). Utilizing Eq. (\ref{eqnonsingtn}), we obtain the physical condition under the OBC \cite{kunst2019transfer},
\begin{align}
    \label{eqappcondition}
    \frac{\sin{\left(N\phi\right)}}{\sin{\left((N-1)\phi\right)}}=q,
\end{align}
with $q=\xi\sqrt{\frac{\mathcal{G}_{wv}}{\mathcal{G}_{vw}}}\mathcal{G}_{vw}$. Further simplifying Eq. (\ref{eqappcondition}) with general complex $\phi=\phi_{R}+i\phi_{I}$ and $\phi_{R}\in[0,2\pi],\phi_{I}\in\mathbb{R}$, we obtain
\begin{align}
    \label{eqappphieq}
    e^{2N\phi_{I}}=\frac{1-qe^{-i\phi}}{1-qe^{i\phi}}e^{2iN\phi_{R}}.
\end{align}
Hence, the solutions of $\phi$ reads
\begin{align}
    \label{eqappobcphisolu}
    \phi_{I}&=\frac{\log{|\mathscr{F}|}}{2N},\nonumber\\
    \phi_{R}&=\frac{2\pi m-\text{arg}(\mathscr{F})}{2N},
\end{align}
where $\mathscr{F}$ labels $\left(1-qe^{-i\phi}\right)/\left(1-qe^{i\phi}\right)$ and $m=1,2,\ldots,N$. Note that $|\mathscr{F}|=1$ leads to $\phi_{I}=0$ if $q$ is real. In general complex-$q$ cases, the explicit $\phi_{I}$ and energy $\varepsilon$ are dependent on $N$ and $\phi_{R}$ through the self-consistent transcendental equations (\ref{eqzequalcos}) and (\ref{eqappobcphisolu}); that is. we can denote $\phi_{I}=c/N$, with $c$ being a finite-value function of $N$ and $\phi_{R}$, and thus, $\phi_{I}$ is quasilinearly dependent on $1/N$.

The solution of an arbitrary eigenstate is given by $\Phi_{n+1}=T^{n}\left(\begin{matrix}1\\0\end{matrix}\right)$, resulting in
\begin{align}
    \label{eqappobceigensolu}
    \beta_{n}&=\frac{\Gamma^{(n-1)/2}}{q\sin{\phi}}\left[\sin{\left((n-1)\phi\right)}-q\sin{\left((n-2)\phi\right)}\right],\nonumber\\
    \alpha_{n}&=\frac{\Gamma^{n/2}}{q\sin{\phi}}\xi\mathcal{G}_{vv}\sin{(n\phi)},
\end{align}
with $n=1,2,\ldots,N$. Note that $\beta_{N+1}=0$ is exactly the physical condition Eq. (\ref{eqappcondition}) and $\alpha_{N}=\tau$. Consequently, an eigenstate corresponding to energy $\varepsilon$ reads
\begin{align}
    \label{eqappobceigenstate}
    \Psi_{n}=\Gamma^{n/2}\left[\mathscr{A}_{L}(\phi)e^{in\phi_{R}}e^{-n\phi_{I}}+\mathscr{A}_{R}(\phi)e^{-in\phi_{R}}e^{n\phi_{I}}\right],
\end{align}
where
\begin{align}
    \label{eqappobcstatecoff}
    \mathscr{A}_{L}(\phi)&=\frac{1}{2iq\sin{\phi}}\left[\xi\mathcal{G}_{vv}v+\left(e^{-i\phi}-qe^{-2i\phi}\right)\frac{w}{\sqrt{\Gamma}}\right],\nonumber\\
    \mathscr{A}_{R}(\phi)&=\frac{1}{2iq\sin{\phi}}\left[-\xi\mathcal{G}_{vv}v+\left(-e^{i\phi}+qe^{2i\phi}\right)\frac{w}{\sqrt{\Gamma}}\right].
\end{align}

\section{The solutions of nonsingular cases with the boundary impurity}
\label{appendixc}
According to the boundary equation (\ref{eqgeneralbeq}) with the boundary impurity, the eigenvalues of the $2\times2$ matrix $KT^{N}$ must be $1$, and 
\begin{align}
    \label{eqappdetktn}
    \det{\left(KT^{N}\right)}=\det{\left(K\right)}\left(\det{T}\right)^{N}=\frac{\gamma_{R}}{\gamma_{L}}\Gamma^{N},
\end{align} 
such that
\begin{align}
    \label{eqapptrktn1}
    \text{tr}\left(KT^{N}\right)=1+\frac{\gamma_{R}}{\gamma_{L}}\Gamma^{N}=\Gamma^{N/2}\left(\frac{\gamma_{R}}{{\gamma_{L}}}\right)^{1/2}\left[\left(\frac{\gamma_{R}}{{\gamma_{L}}}\Gamma^{N}\right)^{-1/2}+\left(\frac{\gamma_{R}}{{\gamma_{L}}}\Gamma^{N}\right)^{1/2}\right].
\end{align}
On the other hand, utilizing formula (\ref{eqnonsingtn}) for nonsingular $T$, we obtain
\begin{align}
    \label{eqapptrktn2}
    \text{tr}\left(KT^{N}\right)=\Gamma^{N/2}\left[\frac{U_{N-1}(z)}{\sqrt{\Gamma}}\text{tr}\left(KT\right)-U_{N-2}(z)\text{tr}\left(K\right)\right]=\Gamma^{N/2}\left[\frac{\text{tr}\left(KT\right)}{\sqrt{\Gamma}}\frac{\sin{\left(N\phi\right)}}{\sin{\phi}}-\text{tr}\left(K\right)\frac{\sin{\left((N-1)\phi\right)}}{\sin{\phi}}\right].
\end{align}
Thus, we get the condition
\begin{align}
    \label{eqappcond}
    \left(\frac{\gamma_{R}}{{\gamma_{L}}}\right)^{1/2}\left[\left(\frac{\gamma_{R}}{{\gamma_{L}}}\Gamma^{N}\right)^{-1/2}+\left(\frac{\gamma_{R}}{{\gamma_{L}}}\Gamma^{N}\right)^{1/2}\right]\sin{\phi}=\frac{\text{tr}\left(KT\right)}{\sqrt{\Gamma}}\sin{\left(N\phi\right)}-\text{tr}\left(K\right)\sin{\left((N-1)\phi\right)}.
\end{align}
Further simplifying with general complex $\phi=\phi_{R}+i\phi_{I}$ and $\phi_{R}\in[0,2\pi],\phi_{I}\in\mathbb{R}$, we obtain
\begin{align}
    \label{eqappphiequ}
    a(\phi)\left(e^{N\phi_{I}}\right)^{2}+b(\phi)e^{N\phi_{I}}+c(\phi)=0,
\end{align}
where 
\begin{align}
    \label{eqappcoff}
    a(\phi)&=\frac{e^{-iN\phi_{R}}}{2i}\left[-\frac{\text{tr}\left(KT\right)}{\sqrt{\Gamma}}+\text{tr}\left(K\right)e^{i\phi}\right],\nonumber\\
    b(\phi)&=-\left[\left(\frac{\gamma_{R}}{{\gamma_{L}}}\Gamma^{N}\right)^{-1/2}+\left(\frac{\gamma_{R}}{{\gamma_{L}}}\Gamma^{N}\right)^{1/2}\right]\left(\frac{\gamma_{R}}{{\gamma_{L}}}\right)^{1/2}\sin{\phi},\nonumber\\
    c(\phi)&=\frac{e^{iN\phi_{R}}}{2i}\left[\frac{\text{tr}\left(KT\right)}{\sqrt{\Gamma}}-\text{tr}\left(K\right)e^{-i\phi}\right].
\end{align}
The solution of $\phi_{I}$ is
\begin{align}
    \label{eqappphisolu}
    e^{N\phi_{I}}=\frac{1}{2a(\phi)}\left[-b(\phi)\pm\sqrt{b^{2}(\phi)-4a(\phi)c(\phi)}\right],
\end{align}
leading to physical real or complex $\phi$ in general. However, since $a(\phi)$ and $b(\phi)$ are always finite and $b(\phi)$ contains the terms $\Gamma^{\pm N/2}$, the emergent forms $\phi_{I}\sim c/N$ with finite size are prevented by the large $N$.

The solution of an arbitrary eigenstate is given by $\Phi_{n+1}=T^{n}\varphi,\, n=1,2,\ldots,N-1$, with $KT^{N}\varphi=\varphi$, resulting in
\begin{align}
    \label{eqappimpsoluofeigen}
    \Phi_{n+1}=\Gamma^{n/2}\left[\mathcal{A}_{L}(\phi)e^{in\phi_{R}}e^{-n\phi_{I}}+\mathcal{A}_{R}(\phi)e^{-in\phi_{R}}e^{n\phi_{I}}\right],
\end{align}
where 
\begin{align}
    \label{eqappimpsolucoff}
    \mathcal{A}_{L}(\phi)&=\frac{1}{2i\sin{\phi}}\left(\frac{T}{\sqrt{\Gamma}}-e^{-i\phi}\mathbbm{1}\right)\varphi,\nonumber\\
    \mathcal{A}_{R}(\phi)&=\frac{-1}{2i\sin{\phi}}\left(\frac{T}{\sqrt{\Gamma}}-e^{i\phi}\mathbbm{1}\right)\varphi.
\end{align}
Further,
\begin{align}
    \label{eqappimpsolualphabeta}
    \beta_{1}&=\left(K_{R}^{-1}\varphi\right)_{1},\nonumber\\
    \alpha_{n}&=\Gamma^{n/2}\left[\mathcal{A}_{L,2}(\phi)e^{in\phi_{R}}e^{-n\phi_{I}}+\mathcal{A}_{R,2}(\phi)e^{-in\phi_{R}}e^{n\phi_{I}}\right],\quad n=1,2,\ldots,N-1,\nonumber\\
    \beta_{n}&=\Gamma^{(n-1)/2}\left[\mathcal{A}_{L,1}(\phi)e^{i(n-1)\phi_{R}}e^{-(n-1)\phi_{I}}+\mathcal{A}_{R,1}(\phi)e^{-i(n-1)\phi_{R}}e^{(n-1)\phi_{I}}\right],\quad n=2,3,\ldots,N,\nonumber\\
    \alpha_{N}&=\left(K_{R}^{-1}\varphi\right)_{2},  
\end{align}
where $\mathcal{A}_{L/R,i}(\phi),i=1,2$, denotes the $i$th component of the column vector $\mathcal{A}_{L/R}(\phi)$. 
Consequently, an eigenstate corresponding to energy $\varepsilon$ reads
\begin{align}
    \label{eqappimpeigenstate}
    \Psi_{1}^{imp}&=\Gamma^{1/2}\left[\mathcal{A}_{L,2}(\phi)e^{i\phi_{R}}e^{-\phi_{I}}+\mathcal{A}_{R,2}(\phi)e^{-i\phi_{R}}e^{\phi_{I}}\right]v+\left(K_{R}^{-1}\varphi\right)_{1}w,\nonumber\\
    \Psi_{n}^{imp}&=\Gamma^{n/2}\left[\mathcal{B}_{L}(\phi)e^{in\phi_{R}}e^{-n\phi_{I}}+\mathcal{B}_{R}(\phi)e^{-in\phi_{R}}e^{n\phi_{I}}\right],\quad n=2,3,\ldots,N-1,\nonumber\\
    \Psi_{N}^{imp}&=\left(K_{R}^{-1}\varphi\right)_{2}v+\Gamma^{(N-1)/2}\left[\mathcal{A}_{L,1}(\phi)e^{i(N-1)\phi_{R}}e^{-(N-1)\phi_{I}}+\mathcal{A}_{R,1}(\phi)e^{-i(N-1)\phi_{R}}e^{(N-1)\phi_{I}}\right]w,
\end{align}
where 
\begin{align}
    \label{eqappimpeigencoffe}
    \mathcal{B}_{L}(\phi)&=\mathcal{A}_{L,2}(\phi)v+e^{-i\phi}\mathcal{A}_{L,1}(\phi)\frac{w}{\sqrt{\Gamma}},\nonumber\\
    \mathcal{B}_{R}(\phi)&=\mathcal{A}_{R,2}(\phi)v+e^{i\phi}\mathcal{A}_{R,1}(\phi)\frac{w}{\sqrt{\Gamma}}.
\end{align}

\section{Details of the HN model with boundary impurity}
\label{appendixd}
Utilizing the single-particle Schr\"odinger equation in Eq. (\ref{eqhaimp}) with the setting $\psi_{0}\equiv\psi_{N}$ , we obtain the bulk and boundary equations
\begin{align}
    t_{R}\psi_{n-1}+t_{L}\psi_{n+1}&=\varepsilon\psi_{n},\label{eqapphnbulk}\\
    \gamma_{R}\psi_{N}+t_{L}\psi_{2}&=\varepsilon\psi_{1},\label{eqapphnboun1}\\
    t_{R}\psi_{N-1}+\gamma_{L}\psi_{1}&=\varepsilon\psi_{N}\label{eqapphnboun2},
\end{align}
where $n\in\mathscr{B}$. The bulk equation (\ref{eqapphnbulk}) gives ($t_{L}\neq0$)
\begin{align}
    \label{eqapphnimpbulkequ}
    \psi_{n+1}=\frac{\varepsilon}{t_{L}}\psi_{n}-\frac{t_{R}}{t_{L}}\psi_{n-1},
\end{align}
thus leading to the propagating relation (\ref{eqhnimpprop}).

\subsection{The case with a singular transfer matrix}
\label{appendixd1}
The real-space EP, or infernal point, emerges under the OBC in the case with the singular transfer matrix ($t_{R}=0$), where all energies are degenerate at zero but have only one eigenvector \cite{,kunst2019transfer,denner2021,fu2022}. Combining the boundary equations (\ref{eqapphnboun1}) and (\ref{eqapphnboun2}) for $t_{R}=0$, we can deduce the physical condition with the boundary impurity,
\begin{align}
    \label{eqapphnsingcond}
    \left(\begin{matrix}\psi_{2}\\\psi_{1}\end{matrix}\right)
    &=\left(\begin{matrix}\frac{\varepsilon}{t_{L}}&-\frac{\gamma_{R}}{t_{L}}\\
    1&0\end{matrix}\right)\left(\begin{matrix}\psi_{1}\\\psi_{N}\end{matrix}\right)\nonumber\\
    &=\left(\begin{matrix}\frac{\varepsilon}{t_{L}}&-\frac{\gamma_{R}}{t_{L}}\\
        1&0\end{matrix}\right)\left(\begin{matrix}\frac{\varepsilon}{\gamma_{L}}&0\\
            1&0\end{matrix}\right)\left(\begin{matrix}\psi_{N}\\\psi_{N-1}\end{matrix}\right)\nonumber\\
    &=\left(\begin{matrix}\frac{\varepsilon}{t_{L}}&-\frac{\gamma_{R}}{t_{L}}\\
        1&0\end{matrix}\right)\left(\begin{matrix}\frac{t_{L}}{\gamma_{L}}&0\\
            0&1\end{matrix}\right)\left(\begin{matrix}\frac{\varepsilon}{t_{L}}&0\\
                1&0\end{matrix}\right)\left(\begin{matrix}\psi_{N}\\\psi_{N-1}\end{matrix}\right)\nonumber\\
    &=\left(\begin{matrix}\frac{\varepsilon}{t_{L}}&-\frac{\gamma_{R}}{t_{L}}\\
        1&0\end{matrix}\right)\left(\begin{matrix}\frac{t_{L}}{\gamma_{L}}&0\\
            0&1\end{matrix}\right)\left(\begin{matrix}\frac{\varepsilon}{t_{L}}&0\\
                1&0\end{matrix}\right)T^{N-2}\left(\begin{matrix}\psi_{2}\\\psi_{1}\end{matrix}\right)\nonumber\\
    &=\left(\begin{matrix}\frac{\varepsilon}{\gamma_{L}}&-\frac{\gamma_{R}}{t_{L}}\\
        \frac{t_{L}}{\gamma_{L}}&0\end{matrix}\right)T^{N-1}\left(\begin{matrix}\psi_{2}\\\psi_{1}\end{matrix}\right)\nonumber\\
    &\equiv K_{d}T^{N-1}\left(\begin{matrix}\psi_{2}\\\psi_{1}\end{matrix}\right),            
\end{align}
where 
\begin{align}
    T=\left(\begin{matrix}\frac{\varepsilon}{t_{L}}&0\\
        1&0\end{matrix}\right)
\end{align}
is the singular transfer matrix with $\Delta=\text{tr}\left(T\right)=\varepsilon/t_{L}$. This implies that the physical eigenvector of $K_{d}T^{N-1}$ corresponds to an eigenvalue $1$. Because $\det{\left(K_{d}T^{N-1}\right)}=\det{\left(K_{d}\right)}\Gamma^{N-1}=0$, we obtain $\text{tr}\left(K_{d}T^{N-1}\right)=1$; thus,
\begin{align}
    \label{eqapphnimpphycond1}
    \text{tr}\left(K_{d}\Delta^{N-2}T\right)=\Delta^{N-2}\text{tr}\left(K_{d}T\right)=\left(\frac{\varepsilon}{t_{L}}\right)^{N-2}\frac{\varepsilon^{2}-\gamma_{L}\gamma_{R}}{t_{L}\gamma_{L}}=1,
\end{align}
which gives 
\begin{align}
    \label{eqapphnimpphycond2}
    \frac{\varepsilon^{2}}{\gamma_{L}}-\gamma_{R}=t_{L}\left(\frac{t_{L}}{\varepsilon}\right)^{N-2}.
\end{align}
Let $\left(t_{L}/\varepsilon\right)^{N-2}=c$, with $c$ being an undetermined coefficient and thus 
\begin{align}
    \label{eqapphnsingenergy}
    \varepsilon=t_{L}c^{-\frac{1}{N-2}}e^{-i\frac{2\pi m}{N-2}},\quad m=1,2,\ldots,N-2.
\end{align}
Substituting Eq. (\ref{eqapphnsingenergy}) into Eq. (\ref{eqapphnimpphycond2}), we obtain the physical condition for $c$,
\begin{align}
    \label{eqapphnimpphycond3}
    \frac{t_{L}^{2}}{\gamma_{L}}c^{-\frac{2}{N-2}}e^{-i\frac{4\pi m}{N-2}}=t_{L}c+\gamma_{R},
\end{align}
where we will label $c_{m}$ as the solution corresponding to $m$. Applying the propagating relation, we obtain
\begin{align}
    \left(\begin{matrix}\psi_{n+1}\\\psi_{n}\end{matrix}\right)&=T^{n-1}\left(\begin{matrix}\psi_{2}\\\psi_{1}\end{matrix}\right)=\Delta^{n-2}T\left(\begin{matrix}\psi_{2}\\\psi_{1}\end{matrix}\right)=\left(\begin{matrix}\left(\frac{\varepsilon}{t_{L}}\right)^{n-1}\psi_{2}\\\left(\frac{\varepsilon}{t_{L}}\right)^{n-2}\psi_{2}\end{matrix}\right),
\end{align}
with $n\in\mathscr{B}$. Together with the boundary equations (\ref{eqapphnboun1}) and (\ref{eqapphnboun2}) for $t_{R}=0$, we finally obtain the eigenvectors with respect to the energies,
\begin{align}
    \label{eqapphnsingvec}
    \varepsilon_{m}&=t_{L}c_{m}^{-\frac{1}{N-2}}e^{-i\frac{2\pi m}{N-2}},\quad m=1,2,\ldots,N,\nonumber\\
    \psi_{n}^{m}&=c_{m}^{-\frac{n-2}{N-2}}e^{-i\frac{2\pi m}{N-2}(n-2)}\psi_{2}^{m},\quad n=3,4,\ldots,N, \nonumber\\
    \psi_{1}^{m}&=\frac{t_{L}}{\gamma_{L}}c_{m}^{-\frac{1}{N-2}}e^{-i\frac{2\pi m}{N-2}}\psi_{N}^{m}.   
\end{align}

Furthermore, we set $t_{L}=e^{\alpha}, t_{R}=e^{-\alpha}, \gamma_{L}=\mu e^{\alpha}, and \gamma_{R}=\mu e^{-\alpha}$; then the strong nonreciprocity $e^{\alpha}\gg e^{-\alpha}$ corresponds to the singular case ($t_{R}\rightarrow0$). In the strong impurity case ($\mu\gg e^{\pm\alpha}$), the physical condition (\ref{eqapphnimpphycond3}) reduces to 
\begin{align}
    \label{eqapplimitstrongcond}
    -\mu e^{-2\alpha}=c,
\end{align}
which leads to the solution according to Eq. (\ref{eqapphnsingvec}), 
\begin{align}
    \label{eqapplimitstrongvec}
    \varepsilon_{m}&=e^{\alpha}e^{-\left(\kappa_{L}+ik_{m}\right)},\nonumber\\
    \psi_{n}^{(m)}&=e^{-\left(\kappa_{L}+ik_{m}\right)(n-2)}\psi_{2}^{(m)},\nonumber\\
    |\psi_{1}^{(m)}&|\sim|\frac{e^{2\alpha}}{\mu^{2}}|\ll1,
\end{align}
where $n=3,4,\ldots,N$, $\kappa_{L}=(\log{\mu}-2\alpha)/(N-2)$, and $k_{m}=\left((2m+1)\pi\right)/(N-2)$. We next consider the limit case $\gamma_{R}\rightarrow0$, and the condition (\ref{eqapphnimpphycond2}) reduces to 
\begin{align}
    \label{eqapplimitcond}
    \left(\frac{\varepsilon}{t_{L}}\right)^{N}=\frac{\gamma_{L}}{t_{L}}.
\end{align}
Then, the eigenvectors with corresponding energies read
\begin{align}
    \label{eqapplimitvec}
    \varepsilon_{m}&=t_{L}^{\frac{1}{N}(N-1)}\gamma_{L}^{\frac{1}{N}}e^{i\frac{2\pi m}{N}},\nonumber\\
    \psi_{n}^{(m)}&=\left(\frac{\gamma_{L}}{t_{L}}\right)^{\frac{n-1}{N}}e^{i\frac{2\pi m}{N}(n-1)}\psi_{1}^{(m)},   
\end{align}
with $m=1,2,\ldots,N$ and $n=2,3,\ldots,N$, which are similar to the results for the $t_{L}=\gamma_{L}=0$ case in Ref. \cite{molignini2023anomalous}. The weak impurity case ($\mu\ll 1$) corresponds to the $\gamma_{R}\rightarrow0$ limit, and the solutiou (\ref{eqapplimitvec}) becomes
\begin{align}
    \label{eqapplimitweaksolu}
    \varepsilon_{m}&=e^{\alpha}e^{i(k_{m}'+i\kappa_{L}')},\nonumber\\
    \psi_{n}^{(m)}&=e^{i(k_{m}'+i\kappa_{L}')(n-1)}\psi_{1}^{(m)},
\end{align}
where $\kappa_{L}'=-\log{\mu}/N$ and $k_{m}'=2\pi m/N$. Noteworthily, Eqs. (\ref{eqapplimitstrongvec}) and (\ref{eqapplimitweaksolu}) are exactly equivalent to the SF effect solutions in Ref. \cite{li2021impurity}.

\subsection{The case with a nonsingular transfer matrix}
\label{appendixd2}

Through boundary equations (\ref{eqapphnboun1}) and (\ref{eqapphnboun2}), we can imitate Eq. (\ref{eqapphnsingcond}) to derive the physical condition in the current case:
\begin{align}
    \label{eqapphnhsshcond1}
    \left(\begin{matrix}\psi_{2}\\\psi_{1}\end{matrix}\right)
    =KT^{N-1}\left(\begin{matrix}\psi_{2}\\\psi_{1}\end{matrix}\right),    
\end{align}
where 
\begin{align}
    \label{eqapphnhsshkmat}
    K=\left(\begin{matrix}
        \frac{\varepsilon}{\gamma_{L}}&-\frac{\gamma_{R}}{t_{L}}\\
        \frac{t_{L}}{\gamma_{L}}&0
    \end{matrix}\right).
\end{align}
Consequently,
\begin{align}
    \label{eqapphnhsshcond2}
    \text{tr}\left(KT^{N-1}\right)&=1+\det{\left(KT^{N-1}\right)}=1+\frac{\gamma_{R}}{\gamma_{L}}\Gamma^{N-1}.  
\end{align}
On the other hand, utilizing formula Eq.~(\ref{eqnonsingtn}) for nonsingular $T$, we obtain ($z=\cos{\phi}$)
\begin{align}
    \label{eqapphnhsshcond3}
    \text{tr}\left(KT^{N-1}\right)=\Gamma^{\frac{N-1}{2}}\left[\frac{U_{N-2}(z)}{\sqrt{\Gamma}}\text{tr}\left(KT\right)-U_{N-3}(z)\text{tr}\left(K\right)\right].
\end{align}
Together with Eq. (\ref{eqhnhssfenformu}) and after some algebra, we obtain the final form of the physical condition,
\begin{align}
    \label{eqapphnhssfcond4}
    \left(\frac{\gamma_{L}}{t_{L}}\Gamma^{-\frac{N}{2}}+\frac{\gamma_{R}}{t_{R}}\Gamma^{\frac{N}{2}}\right)\sin{\phi}=\sin{\left((N+1)\phi\right)}-\frac{\gamma_{L}\gamma_{R}}{t_{L}t_{R}}\sin{\left((N-1)\phi\right)},
\end{align}
which is exactly the result in Ref. \cite{guo2021exact}. The solutions of $\phi$ in Eq. (\ref{eqapphnhssfcond4}) are real or complex values for different parameters according to Ref. \cite{guo2021exact}, and the imaginary part of the complex solution $\phi=\phi_{R}+i\phi_{I}$ can take the form $\phi_{I}=c/N$, with $c$ being dependent on $N$ and $\phi_{R}$ as before.

The OBC condition corresponding to $\gamma_{L}=\gamma_{R}=0$ reduces Eq. (\ref{eqapphnhssfcond4}) to $\sin{\left((N+1)\phi\right)}=0$, which leads to the solutions $\phi=l\pi/(N+1)\in\left[0,\pi\right]$, with $l=1,2,\ldots,N$. The eigenstate corresponding to energy $\varepsilon=2\sqrt{t_{L}t_{R}}\cos{\phi}$ is deduced as
\begin{align}
    \label{eqapphnobcvec}
    \left(\begin{matrix}\psi_{n+1}\\\psi_{n}
    \end{matrix}\right)=T^{n}\left(\begin{matrix}1\\0
    \end{matrix}\right)=\frac{1}{\sin{\phi}}\left(\begin{matrix}\Gamma^{n/2}\sin{\left((n+1)\phi\right)}\\\Gamma^{(n-1)/2}\sin{\left(n\phi\right)}
    \end{matrix}\right),
\end{align}
and is normalized as $\psi_{n}=\mathcal{N} \Gamma^{n/2}\sin{\left(n\phi\right)}$, with $\mathcal{N}$ being the normalization coefficient, which is exactly the bulk eigenstate of the pure finite-size skin effect.

\section{Discussion of the quasilinear dependence of the pure SF effect}
\label{appendixadd}

\subsection{HN model}
\label{appendixadd1}
Consider the physical condition (\ref{eqhnimpphycond3}) for the HN model with singular a transfer matrix. Assume $c_{m}\sim0$; then the physical condition reduces to 
\begin{align}
    \label{appeqhnzerolimitcon}
    \frac{t_{L}^{2}}{\gamma_{L}}c_{m}^{-\frac{2}{N-2}}e^{-i\frac{4\pi m}{N-2}}=\gamma_{R},
\end{align}
resulting in 
\begin{align}
    \label{apphnzerosolution}
    &c_{m}^{-\frac{2}{N-2}}=\frac{\gamma_{L}\gamma_{R}}{t_{L}^{2}}e^{i\frac{4\pi m}{N-2}},\nonumber\\
    &c_{m}^{-1}=\left(\frac{\gamma_{L}\gamma_{R}}{t_{L}^{2}}\right)^{\frac{N-2}{2}},\nonumber\\
    &c_{m}=\left(\frac{t_{L}^{2}}{\gamma_{L}\gamma_{R}}\right)^{\frac{N-2}{2}}.
\end{align}
If $|\frac{t_{L}^{2}}{\gamma_{L}\gamma_{R}}|<1$ and $N$ is large enough (e.g., in the thermodynamic limit), $c_{m}\sim0$ holds, and the solution (\ref{apphnzerosolution}) is a valid solution of Eq. (\ref{eqhnimpphycond3}) with $\text{Re}\left(\log{c_{m}}\right)$ linearly dependent on $N-2$. The localization length
\begin{align}
    \label{apphnzerolimitlength}
    \xi_{m}=\frac{N-2}{|\text{Re}\left(\log{c_{m}}\right)|}\sim\frac{N-2}{\left|\text{Re}\left(\frac{N-2}{2}\log{\left(\frac{t_{L}^{2}}{\gamma_{L}\gamma_{R}}\right)}\right)\right|}=\frac{1}{\left|\frac{1}{2}\log{\left(\frac{t_{L}^{2}}{\gamma_{L}\gamma_{R}}\right)}\right|}
\end{align}
is independent of $N$, which is the emergence of the pure skin effect. According to Eq. (\ref{eqhnsingvec}), the eigenstates in the bulk are $\psi_{n}\sim c_{m}^{-n/(N-2)}=\left(\sqrt{\frac{\gamma_{L}\gamma_{R}}{t_{L}^{2}}}\right)^{n}$, with the phase $e^{-i\frac{2\pi m}{N-2}n}$ for each $m$, resulting in the circle GBZ with radius $\left|\sqrt{\frac{\gamma_{L}\gamma_{R}}{t_{L}^{2}}}\right|$. However, if $|\frac{t_{L}^{2}}{\gamma_{L}\gamma_{R}}|\geq1$ or $N$ is finite, $\text{Re}\left(\log{c_{m}}\right)$ is not linearly dependent on $N-2$, and $\xi_{m}$ is quasilinearly dependent on $N$ without explicit expression in general.

\subsection{NH-SSH model}
\label{appendixadd2}
Consider the physical condition (\ref{eqnsshsingcond}) of the NH-SSH model with a singular transfer matrix. Assume $c_{m}\sim0$; then the physical condition reduces to
\begin{align}
    \label{appeqnsshaerolomitcon}
    t_{2}(1-\gamma_{L}\gamma_{R})=-2\gamma c_{m}^{-\frac{1}{N}}e^{-i\frac{2\pi m}{N}},
\end{align}
resulting in 
\begin{align}
    \label{appnsshzerosolution}
    &\frac{t_{2}(\gamma_{L}\gamma_{R}-1)}{2\gamma}e^{i\frac{2\pi m}{N}}=c_{m}^{-\frac{1}{N}},\nonumber\\
    &c_{m}^{-1}=\left(\frac{t_{2}(\gamma_{L}\gamma_{R}-1)}{2\gamma}\right)^{N},\nonumber\\
    &c_{m}=\left(\frac{2\gamma}{t_{2}(\gamma_{L}\gamma_{R}-1)}\right)^{N}.
\end{align}
If $|\frac{2\gamma}{t_{2}(\gamma_{L}\gamma_{R}-1)}|<1$ and $N$ is large enough (e.g., in the thermodynamic limit), $c_{m}\sim0$ holds, and the solution (\ref{appnsshzerosolution}) is a valid solution of Eq. (\ref{eqnsshsingcond}) with $\text{Re}\left(\log{c_{m}}\right)$ linearly dependent on $N$. The localization length
\begin{align}
    \label{appnsshpurezerolimitlength}
    \xi_{m}=\frac{N}{|\text{Re}\left(\log{c_{m}}\right)|}\sim\frac{N}{\left|\text{Re}\left(N\log{\left(\frac{2\gamma}{t_{2}(\gamma_{L}\gamma_{R}-1)}\right)}\right)\right|}=\frac{1}{\left|\log{\left(\frac{2\gamma}{t_{2}(\gamma_{L}\gamma_{R}-1)}\right)}\right|}
\end{align}
is independent of $N$, which is the emergence of the pure skin effect. According to Eq. (\ref{eqnsshsingsolu}), the eigenstates in the bulk are $\Psi_{n}\sim c_{m}^{-n/N}=\left(\frac{t_{2}(\gamma_{L}\gamma_{R}-1)}{2\gamma}\right)^{n}$, with the phase $e^{-i\frac{2\pi m}{N}n}$ for each $m$, resulting in the circle GBZ with radius $\left|\frac{t_{2}(\gamma_{L}\gamma_{R}-1)}{2\gamma}\right|$. However, if $|\frac{2\gamma}{t_{2}(\gamma_{L}\gamma_{R}-1)}|\geq1$ or $N$ is finite, $\text{Re}\left(\log{c_{m}}\right)$ is not linearly dependent on $N$, and $\xi_{m}$ is quasilinearly dependent on $N$ without explicit expression in general.

\section{The emergent hybrid SFS modes of the NH-SSH model with the boundary impurity}
\label{appendixe}
\begin{figure*} 
    \centering 
    \subfigure[]{\includegraphics[width=4cm, height=2.8cm]{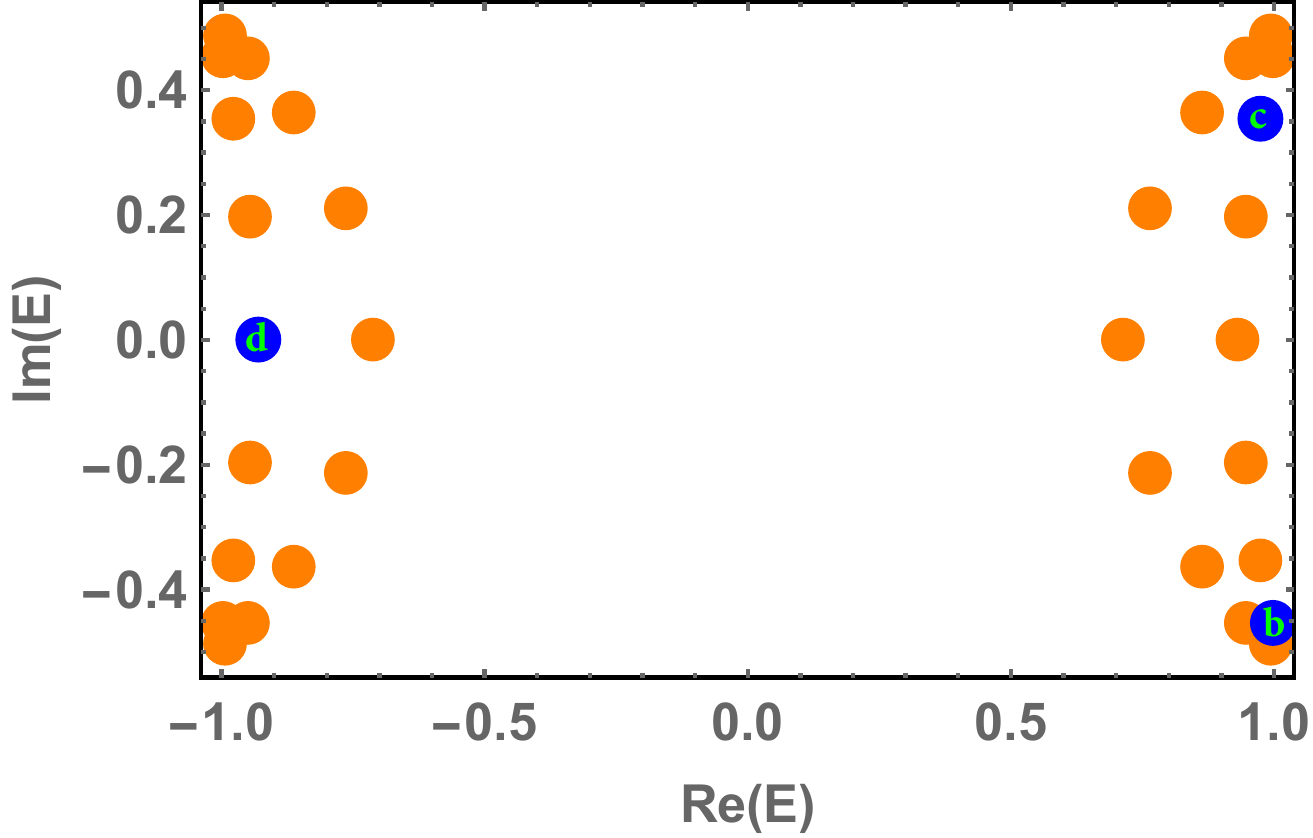}}
    \subfigure[]{\includegraphics[width=3.9cm, height=2.8cm]{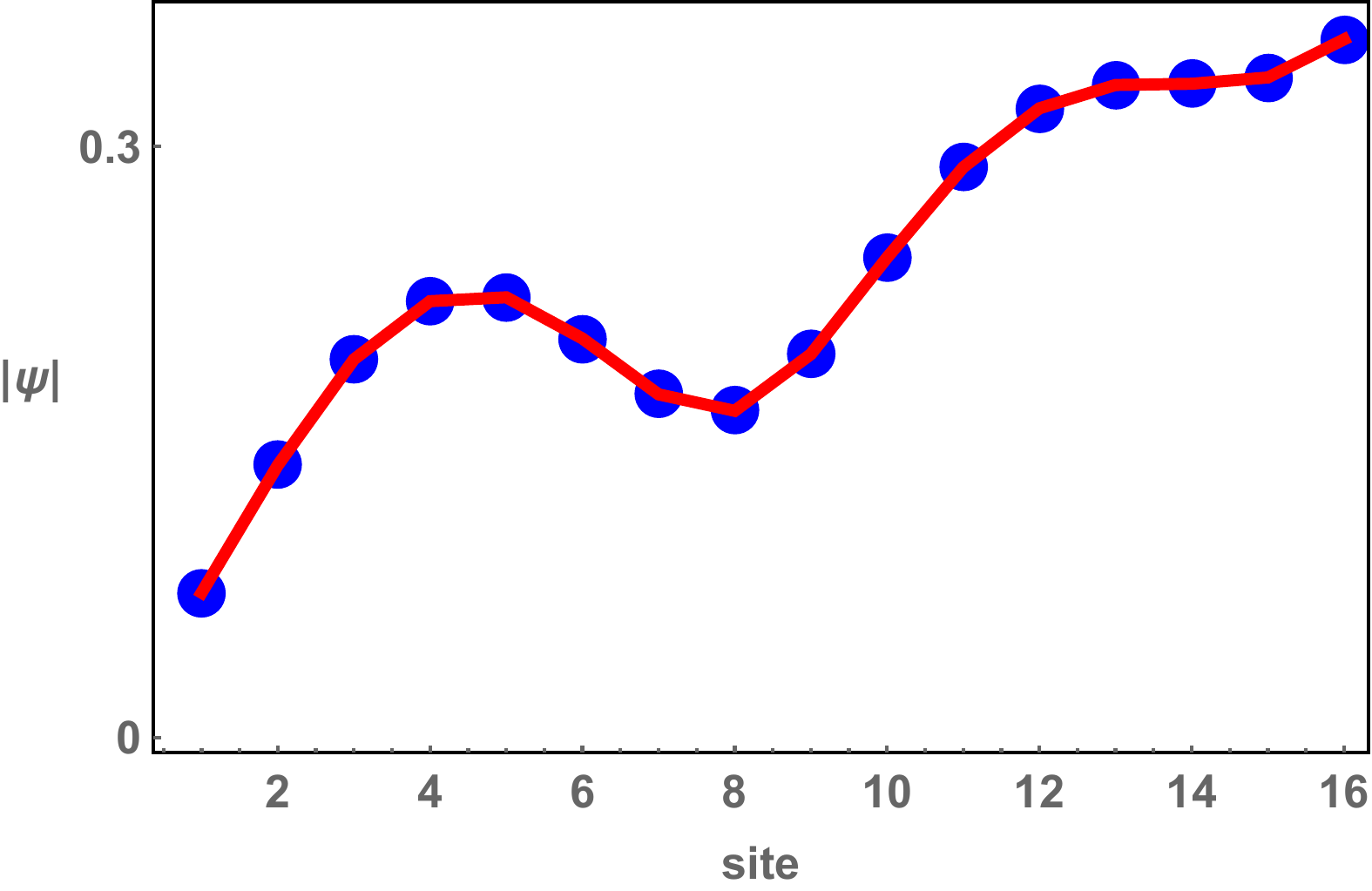}}
    \subfigure[]{\includegraphics[width=3.6cm, height=2.8cm]{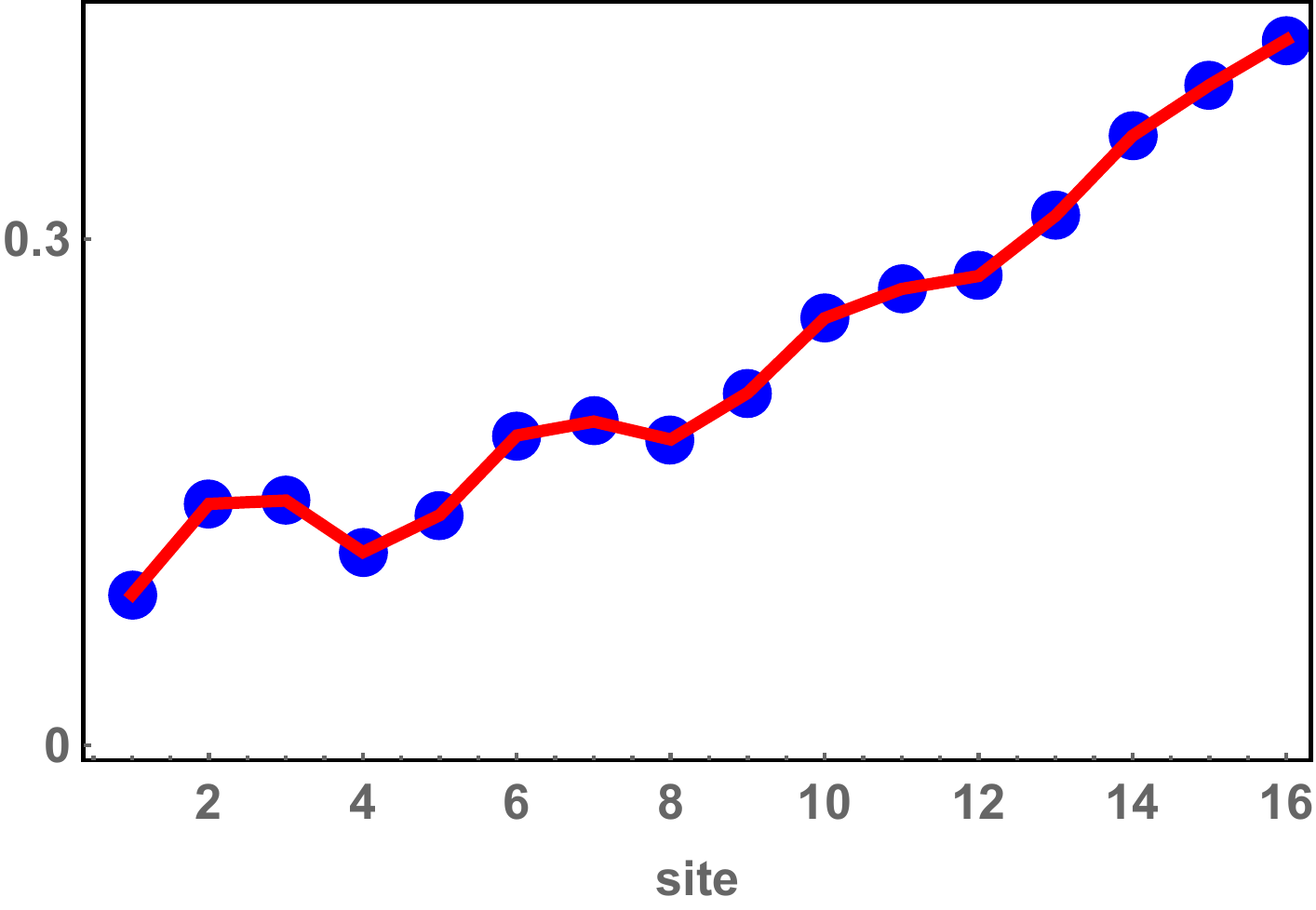}}
    \subfigure[]{\includegraphics[width=3.6cm, height=2.8cm]{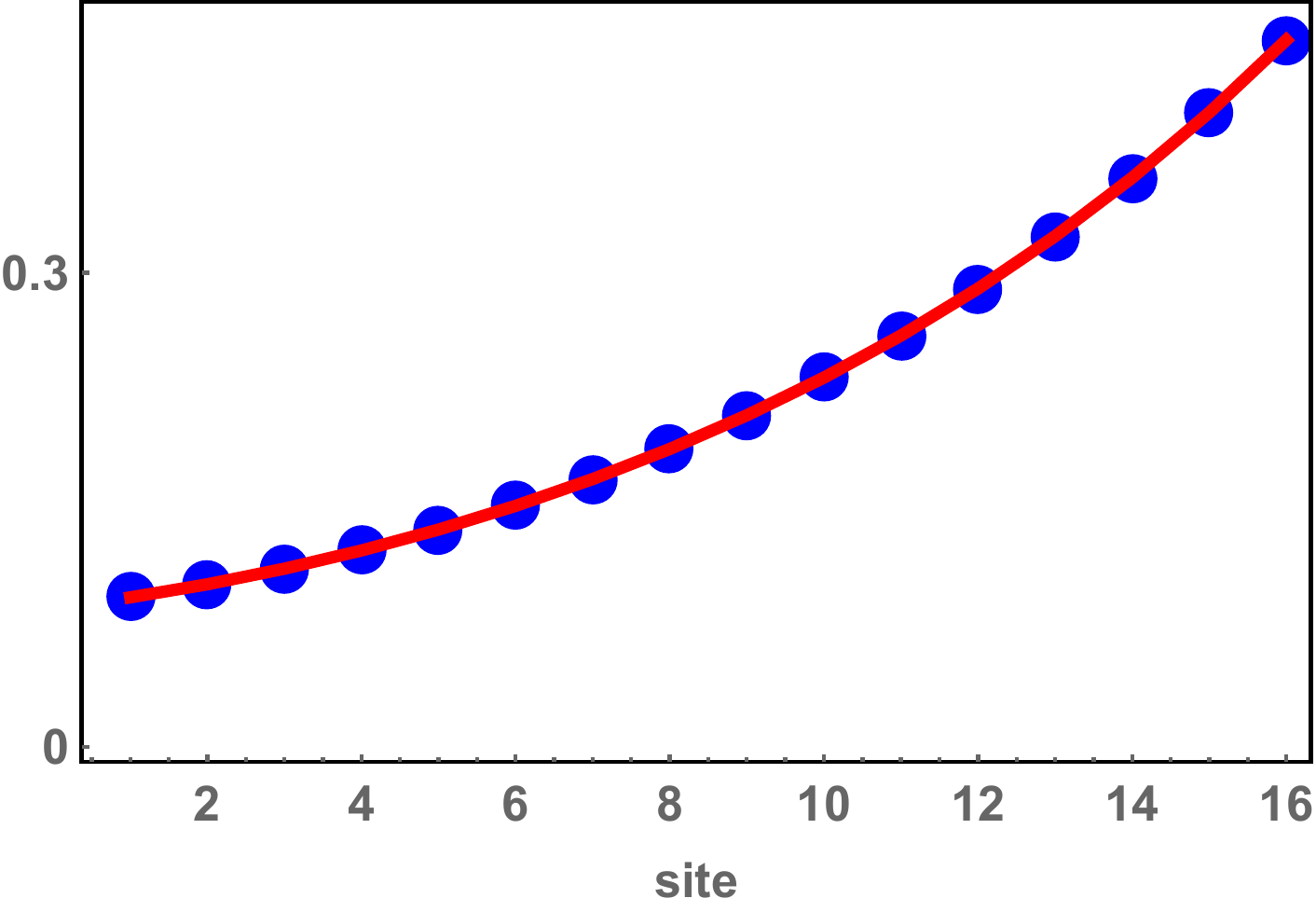}}
    \caption{(a) The energy spectrum of the Hamiltonian $\bar{\mathcal{H}}_{imp}$ for parameters $t_{1}=0.1,t_{2}=1,\gamma=0.5,\gamma_{L}=0.3,\gamma_{R}=1.2$, and $N=16$. (b)-(c) The analytic (red line) and numerical (blue dots) results of the pure SF modes corresponding to the blue dots in (a).}
    \label{Fig-nsshimphssf} 
\end{figure*}

We again perform the generalized gauge transformation in Sec. \ref{section4b} on $\hat{\mathcal{H}}_{nssh}$ together with the boundary impurity, and obtain the target Hamiltonian $\bar{\mathcal{H}}_{imp}$, the Hamiltonian $\bar{\mathcal{H}}$ in Sec. \ref{section4b}, together with boundary impurity strengths $\bar{\gamma}_{L}=r^{-N}\gamma_{L}$ and $\bar{\gamma}_{R}=r^{N}\gamma_{R}$. The solutions of the emergent pure SF modes are Eq. (\ref{eqappimpeigenstate}) with $\bar{\Gamma}=1$ and $\varphi=\bar{K}_{R}\Phi_{1}$, where
\begin{align}
    \Phi_{1}=\left(\begin{matrix}
        \psi_{1}\\\psi_{2N}
    \end{matrix}\right),\quad
    \bar{K}_{R}=\left(\begin{matrix}
        1&0\\
        0&\bar{\gamma}_{R}
    \end{matrix}\right),
\end{align}  
and $\psi$ is the corresponding numerical eigenvector. The energy spectrum and a comparison between analytic and numerical results of the pure SF modes are illustrated in Fig. \ref{Fig-nsshimphssf} for selected parameters, which expectedly match each other. Therefore, the pure SF effect of $\bar{\mathcal{H}}_{imp}$ implies the finite-size hybrid SFS effect of $\hat{\mathcal{H}}_{nssh}$ with the boundary impurity.

\end{widetext}

\bibliography{reference}
\end{document}